\PassOptionsToPackage{svgnames,table}{xcolor}
\documentclass[11pt]{article}
\pdfoutput=1 %for arxiv
\usepackage{palatino,microtype}
\usepackage[margin=15pt,font=small,labelfont={bf}]{caption}
\usepackage[margin=1in]{geometry}
\usepackage[english]{babel}
\usepackage[utf8]{inputenc}%[utf8x]
\usepackage[compact]{titlesec}
\usepackage{cmap}
\usepackage[T1]{fontenc}
\usepackage{bm}
\usepackage[colorinlistoftodos]{todonotes}
\usepackage{booktabs}
\usepackage{mathtools}
\usepackage{amsmath}
\usepackage{amssymb}
\usepackage{amsthm}
\usepackage{float}
\usepackage{graphics}
\usepackage[numbers]{natbib}
\usepackage{hyperref}
\hypersetup{colorlinks={true},urlcolor={blue},linkcolor={DarkBlue},citecolor=[named]{DarkGreen},linktoc=all}
\usepackage{colortbl}
\colorlet{myred}{red!25}
\colorlet{myblue}{blue!25}
\colorlet{mygreen}{green!25}

\newcommand{\BibTeX}{\rm B\kern-.05em{\sc i\kern-.025em b}\kern-.08em\TeX}

\usepackage[labelfont={normalfont,bf},textfont=it]{caption}
\usepackage{subcaption}

\usepackage{nicefrac}
\usepackage{pifont}

\usepackage{enumitem}
\usepackage{apxproof}
\usepackage{array,multirow,graphicx,bigdelim}
\input{insbox}%%%%%%%%%%%%%% TeX macro,

\usepackage{tikz}
\usepackage{tikz-dimline}
\usepackage{tikz-cd}
\usetikzlibrary{fit,calc,math,shapes,shapes.multipart,decorations.text,arrows,decorations.markings,decorations.pathmorphing,shapes.geometric,positioning,decorations.pathreplacing}

\tikzset{snake it/.style={decorate, decoration=snake}}

\tikzset{
    leading_agent/.style={circle, draw={rgb, 255:red, 93; green, 166; blue, 13}, 
                          fill={rgb, 255:red, 232; green, 254; blue, 212}, 
                          line width=0.75pt, minimum size=6mm, inner sep=0pt}, % Leading agent node style
    agent/.style={circle, draw=black, fill=lightgray!30, line width=0.75pt, minimum size=6mm, inner sep=0pt}, % Agent node style
    etc/.style={draw=none, minimum size=6mm, inner sep=0pt},
    envy/.style={-{Triangle}, line width=1pt}, 
    champ/.style={-{Triangle}, bend left=20, color={rgb, 255:red, 208; green, 2; blue, 27}, line width=1pt}, 
    best/.style={-{Triangle}, bend left=20, color=teal, line width=1pt},
    every loop/.style={min distance=12mm}
}

\colorlet{mygray}{gray!40}

\makeatletter
\let\oldnl\nl% Store \nl in \oldnl
\newcommand{\nonl}{\renewcommand{\nl}{\let\nl\oldnl}}% Remove line number for one line
\makeatother

\usepackage{thmtools,thm-restate}
\usepackage[capitalise,nameinlink,noabbrev]{cleveref}

\newtheorem{definition}{Definition}%[section]
\newtheorem{lemma}{Lemma}%[section]
\newtheorem{theorem}{Theorem}
\newtheorem{corollary}{Corollary}
\newtheorem{proposition}{Proposition}
\newtheorem{invariant}{Invariant}
\newtheorem{conjecture}{Conjecture}

\Crefname{claim}{Claim}{Claims}
\Crefname{corollary}{Corollary}{Corollaries}
\Crefname{definition}{Definition}{Definitions}
\Crefname{example}{Example}{Examples}
\Crefname{lemma}{Lemma}{Lemmas}
\Crefname{property}{Property}{Properties}
\Crefname{proposition}{Proposition}{Propositions}
\Crefname{remark}{Remark}{Remarks}
\Crefname{theorem}{Theorem}{Theorems}

\newcommand{\M}{\mathcal{M}} %Good set
\newcommand{\N}{\mathcal{N}} %Agent set
\newcommand{\V}{\mathcal{V}} %Valuation tuple
\newcommand{\R}{\mathbb{R}_{{\ge}0}} %Non negerive Real
\newcommand{\A}{\mathbb{A}}
\newcommand{\B}{\mathbb{B}}
\newcommand{\C}{\mathbb{C}}

\usepackage[linesnumbered,ruled,vlined]{algorithm2e}  % For the algorithm environment

\theoremstyle{remark}

\newcommand{\myBox}[3]{%
  \renewcommand{\arraystretch}{1.3}% Adjust the value for more/less padding
  \begin{array}{|c|} 
    \hline 
    #1 \\
    \hline 
    #2 \\ 
    \hline  
    \multicolumn{1}{c}{} \\[-1.2em] 
    \multicolumn{1}{c}{#3} 
  \end{array}%
  \renewcommand{\arraystretch}{1}% Reset arraystretch back to default
}

\newcommand{\myBoxtwo}[2]{%
    \renewcommand{\arraystretch}{1.3}% Adjust the value for more/less padding
    \begin{array}{|c|} 
        \hline 
        \\ 
        {\raisebox{0.7em}[0cm][0cm]{\ensuremath{#1}}} \\ 
        \hline 
        \multicolumn{1}{c}{} \\[-1.2em] 
        \multicolumn{1}{c}{#2} 
    \end{array}%
    \renewcommand{\arraystretch}{1}% Adjust the value for more/less padding
}

\newcommand{\Allocation}[6]{%
  \myBox{#1}{#2}{a_1} \hspace{2em}%
  \myBox{#3}{#4}{b_1} \hspace{2em}%
  \myBox{#5}{#6}{c_1}%
}

\newcommand{\Allocationfivetwo}[5]{%
  \myBox{#1}{#2}{a_1} \hspace{2em}%
  \myBoxtwo{#3}{b_1} \hspace{2em}%
  \myBox{#4}{#5}{c_1}%
}

\newcommand{\Allocationfiveone}[5]{%
  \myBoxtwo{#1}{a_1} \hspace{2em}%
  \myBox{#2}{#3}{b_1} \hspace{2em}%
  \myBox{#4}{#5}{c_1}%
}

\newcommand{\mes}{\mathsf{MES}}

\let\displaystyle\textstyle

\title{EFX Exists for Three Types of Agents}
\author{
	\begin{tabular}{m{0.12\textwidth}m{0.12\textwidth}m{0.12\textwidth}m{0.12\textwidth}%m{0.12\textwidth}m{0.12\textwidth}
 }
    \multicolumn{2}{c}{\textbf{Vishwa Prakash HV}} & \multicolumn{2}{c}{\textbf{Pratik Ghosal}}
        \\
        \multicolumn{2}{c}{Chennai Mathematical Institute} & \multicolumn{2}{c}{Indian Institute of Technology Palakkad} %& \multicolumn{2}{c}{Indian Institute of Technology Delhi}
        \\ 
		    \multicolumn{2}{c}{\href{mailto:vishwa@cmi.ac.in}{\small{\texttt{vishwa@cmi.ac.in}}}} &\multicolumn{2}{c}{\href{mailto:pratik@iitpkd.ac.in}{\small{\texttt{pratik@iitpkd.ac.in}}}}
        \\
        &&&\\
		\multicolumn{2}{c}{\textbf{Prajakta Nimbhorkar}} & \multicolumn{2}{c}{\textbf{Nithin Varma}} 
        \\
    \multicolumn{2}{c}{Chennai Mathematical Institute} & 
    \multicolumn{2}{c}{\begin{tabular}{c}
        University of Cologne, Germany\\
        Max Planck Institute for Informatics, Germany
    \end{tabular}}
    \\
		\multicolumn{2}{c}{\href{mailto:prajakta@cmi.ac.in}{\small{\texttt{prajakta@cmi.ac.in}}}} & \multicolumn{2}{c}{\href{mailto:nmahendr@uni-koeln.de}{\small{\texttt{nmahendr@uni-koeln.de}}}}
	\end{tabular}
}

\DeclareMathOperator*{\argmax}{arg\,max}
\DeclareMathOperator*{\argmin}{arg\,min}

\date{}

\begin{document}

\maketitle 
 
%%%%%%%%%%%%%%%%%%%%%%%%%%%%%%%%%%%%%%%%%%%%%%%%%%%%%%%%%%%%%%%%%%%%%%%%

\begin{abstract}\label{sec:abstract}
    We study the problem of finding an envy-free allocation of indivisible goods among agents with additive valuations. We focus on the fairness notion of \textit{envy-freeness up to any good} (EFX). A central open question in fair division is whether EFX allocations always exist for any number of agents. While EFX has been established for three agents \cite{efx_3} and for any number of agents with at most two distinct valuations \cite{mahara}, its existence in more general settings remains open.

    In this paper, we make significant progress by proving that EFX allocations exist for any number of agents when there are at most three distinct additive valuations. This result simultaneously generalizes both the three-agent case and the two-type case, settling an open question in the field (see \cite{mahara}).
\end{abstract}

\section{Introduction}\label{sec:intro}

Fair allocation of indivisible resources is a fundamental problem at the intersection of economics and computer science. The objective here is to distribute a set of indivisible goods among multiple agents with different preferences in a fair manner. This problem arises in a wide range of practical scenarios, from assigning slots and assets to distributing aid and shared resources. One of the most natural and widely studied fairness notions is envy-freeness (EF) \cite{foley1967resource}, which ensures that no agent prefers another's allocation over their own.

However, envy-free allocations do not always exist, making exact envy-freeness unattainable in many cases. Moreover, even when an EF allocation exists, finding one is computationally intractable. 
To overcome these challenges, \cite{efx_no1} introduced envy-freeness up to any good (EFX), a minimal and natural relaxation that has gained prominence since then. Under EFX, an agent may envy another, but this envy can always be eliminated by removing any one good from the envied agent's bundle. EFX is widely regarded as the strongest known fairness criterion that remains potentially achievable for indivisible goods. Quoting~\cite{CaragiannisGH19},
\begin{quote}
    ``\emph{Arguably, EFX is the best fairness analog of envy-freeness for indivisible items.}''
\end{quote}
Yet, despite its intuitive appeal, the existence of EFX allocations under general conditions has remained an enigmatic open problem. \cite{procacciaTechnicalperspective20} remarks that:
\begin{quote}
    ``\emph{This fundamental and deceptively accessible question is open. In my view, it is the successor of envy-free cake cutting as fair division’s biggest problem.}''
\end{quote}
%---even in many restricted settings, where only partial results are known for specific cases \cite{efx_no1, efx_3, mahara, pr20, ghosal2023efxexistsagentstypes, BergerCFF22}.          

To discuss EFX allocations in more detail, we recall the standard framework for discrete fair allocation. The setting consists of a set of \( n \) agents, \( \N = \{a_1, a_2, \dots, a_n\} \), and a set of \( m \) \emph{indivisible} goods, \( \M = \{g_1, g_2, \dots, g_m\} \). Each agent \( a_i \in \N \) has a valuation function \( v_i: 2^\M \to \mathbb{R}_{\geq 0} \), which quantifies the utility that \(a_i\) derives from any subset of goods. We focus on \emph{additive valuations}, meaning the value of a subset of goods is simply the sum of the values of its individual items. Formally, for any subset \( S \subseteq \M \), the valuation function satisfies  \( v_i(S) = \sum_{g \in S} v_i(\{g\}) \). An allocation \( X = \langle X_{a_1}, X_{a_2}, \dots, X_{a_n} \rangle \) is a partition of \( \M \) into \( n \) disjoint subsets, called \emph{bundles}, where each agent \( a_i \) receives the bundle \( X_{a_i} \).

\paragraph{EFX Allocations:} Given an allocation \(X\), an agent \(a_i\) \emph{envies} an agent \(a_j\) if \(v_i(X_{a_i}) < v_i(X_{a_j})\). This envy is said to be \emph{weak} if it disappears after removing any one  good from \(X_{a_j}\); otherwise, it is considered \emph{strong}. An allocation is \emph{envy-free up to any good} (EFX) if no agent has a strong envy towards any other agent. The question of whether EFX allocations always exist for \(n\) agents with arbitrary additive valuations remains open. %However, over the years, several partial results have been established. 
Following its introduction by \cite{efx_no1}, \cite{pr20} proved the existence of EFX allocations in two specific cases: \((i)\) when there are only two agents and \((ii)\) for any number of agents, provided that all agents have identical valuations. \cite{mahara} generalized this result by proving that an EFX allocation always exists when agents belong to one of two types, where agents of the same type have identical additive valuations. In a breakthrough result, \cite{efx_3} established that EFX allocations always exist for instances with three agents. 
{Whether complete EFX allocations exist in significantly more general settings, i.e., either for more number of agents or more than two distinct valuations remains wide open (see, e.g., \cite{mahara}).}
%More recently, \cite{ghosal2023efxexistsagentstypes} further generalized this result to \(n\) agents when at least \(n-2\) of them have identical valuations.

\subsection{Our Contributions}
{In this paper, we significantly advance the state-of-the-art by showing the existence of EFX allocations for instances with three types of agents. Specifically, we provide a constructive proof of the following theorem thereby solving an open question highlighted by~\cite{mahara}.} 

\begin{theorem}\label{thm:efx3}
    An EFX allocation always exists for \(n\) agents when there are at most three types of agents, where agents of the same type have identical additive valuations.
\end{theorem}

Our result generalizes and unifies both the
three-agent case~\cite{efx_3} and the two-types case~\cite{mahara} making a substantial contribution to the field. 

Our proof, in addition to using some of the existing techniques in the field, involves several new ideas. We believe that these ideas can be helpful in making further progress on EFX. We outline some of these ideas at a high level in what follows.
\noindent\paragraph{Progress Measure: }Our proof is algorithmic in nature and relies on a combination of Pareto improvements and a novel potential function \(\phi\), specifically designed for the three-types setting. The potential function is the value of the minimum-valued bundle given to any agent in a fixed group (see Section~\ref{subsec:progress} for a formal definition). %It generalizes the lexicographic potential function introduced by \cite{efx_no1}. 
We start with a trivial partial EFX allocation and iteratively update it. Given any partial EFX allocation, our algorithm constructs a new (possibly partial) EFX allocation with a strictly larger potential. This ensures that the process always terminates with a complete EFX allocation.

\noindent\paragraph{Pseudo-Cycles:} 
All earlier work on EFX allocations have crucially exploited the structural patterns arising from various envy relationships among agents, in order to reallocate goods and improve allocations. Examples of these include envy cycles \cite{lipton} and champion cycles \cite{chaudhury2021little}.
%Envy cycles \cite{lipton} and champion cycles \cite{chaudhury2021little} have been used in literature. 
In this work, we introduce a {novel} notion of \emph{pseudo-cycles} (Section~\ref{subsec:pseudo_cycles}) that
captures certain envy relations among \emph{groups of agents}. 
While champion cycles and envy cycles can guide the reallocation of goods and result in Pareto improvement, pseudo-cycles enable us to reallocate goods and increase our potential function, even by possibly assigning lower valued bundles to some agents.
%While elimination of envy cycles and champion cycles always results in a Pareto improvement, the same is not necessarily true for pseudo-cycles. Nevertheless, we show that eliminating pseudo-cycles leads to an increase in our potential function. 
%We believe that the techniques developed here could also provide insights for handling EFX allocations in the more general k-type setting.

\noindent\paragraph{Making Progress Through Regress, and Virtual Bundles:} Our proof goes through an involved case analysis of various configurations of the {directed graph of envies among agents (envy graph).} %envy graph. 
In one of the cases, to get a new EFX allocation with a larger value of the potential function $\phi$, we need to iteratively deallocate some goods as an intermediate step. {Additionally, we introduce the idea of {\em virtual bundles}, where an agent, say $p$, considers a subset of agents and hypothetically assigns smaller virtual bundles to those agents, such that this hypothetical allocation is acceptable to $p$.}

For this case, we design an iterative algorithm called the \textsc{Iterated-Competition} algorithm (see Algorithm~\ref{alg:multi-competition} in Section \ref{sec:competition}). In this algorithm, we use three hierarchical potential functions to track progress. In each iteration of the algorithm, one of the potential functions in the hierarchy improves, while keeping the potential function(s) that are higher up in the hierarchy fixed. 
The potential functions in the decreasing order of precedence are (1) $\phi$, described earlier, (2) the number of unallocated goods, and (3) the number of virtual bundles.
%Additionally, we introduce the idea of {\em virtual bundles}, where an agent, say $p$, creates virtual copies of the bundles of some other agents and assigns them a smaller bundle in that local copy such that any of these smaller bundles are {\em acceptable} to $p$.
%The first one in the hierarchy is $\phi$, described earlier. 
As long as we cannot find an EFX allocation with an improved value of $\phi$, the second potential function is increased while keeping $\phi$ the same. 
 If neither of the two potential functions described above can be increased, then the third potential function increases, keeping the other two fixed.

%Our proof involves a detailed case-by-case analysis of various configurations of the envy graph. One of the most technically intricate cases arises when a particular agent does not envy anyone. To handle this scenario, we introduce a novel iterative algorithm called the \textsc{Iterated-Competition} algorithm (see Algorithm~\ref{alg:multi-competition}). This algorithm tracks progress using three hierarchical potential functions. When direct improvements in the primary potential function are not possible, the algorithm carefully shrinks the allocation while maintaining the potential value. If even this approach reaches a deadlock, the algorithm iteratively applies a technique we call \emph{virtual deletion}, creating \emph{virtual bundles}. This allows progress while preserving the allocation. Ultimately, we prove that these steps always lead to a potential function improvement, ensuring termination.

\subsection{Related Work}
The literature on the fair division of indivisible goods is extensive. This section highlights the works most relevant to our study - EFX allocations of goods; for further details, we refer the reader to a recent survey by \cite{amanatidis2023fair}.

In addition to the studies on complete EFX allocations discussed above, several works have focused on \emph{almost} complete EFX allocations, where the objective is to minimize the number of unallocated goods.  A key result by \cite{chaudhury2021little} showed that an EFX allocation always exists with at most \(n-1\) unallocated goods, such that no agent envies the unallocated set. Subsequent work by \cite{BergerCFF22} and \cite{Mahara21} improved this bound, {using much more sophisticated techniques}, proving that EFX allocations exist with at most \(n-2\) unallocated goods. Additionally, they proved that for four agents, an EFX allocation can always be found with at most one unallocated good. More recently, \cite{ghosal2023efxexistsagentstypes} generalized this result, showing that for any number of agents with at most \(k\) distinct valuations, an EFX allocation exists with at most \(k-2\) unallocated goods. {In particular, the result of \cite{ghosal2023efxexistsagentstypes} implies that one can obtain a partial EFX allocation with at most  one unallocated good for the setting with three distinct types of agents. Our result directly improves upon this by obtaining a complete EFX allocation in this setting. In this context, we would like to stress the fact that obtaining complete EFX allocations is significantly more challenging than obtaining partial EFX allocations. }

An interesting constraint is to restrict the number of possible values agents can have for the goods. \cite{amanatidis2021maximum} showed that EFX allocations exist for bi-valued instances, where each agent assigns one of two possible values to each good. \cite{garg2023computing} extended this result by showing that Pareto optimal EFX allocations always exist for bi-valued instances. 

There have been several more studies on approximate notions of EFX allocations. An allocation is said to be \(\alpha\)-EFX, if for every pair of agents \(a_i\) and \(a_j\), agent \(a_i\) finds her bundle \(X_{a_i}\) to be at least as valuable as \(\alpha\cdot v_i(X_{a_j}\setminus g)\) for any \(g\in X_{a_j}\). \cite{pr20} showed the existence of \(\frac{1}{2}\)-EFX allocations. Following this, \cite{AMANATIDIS202094} showed that \(0.618\)-EFX allocations always exists. In a recent work, ~\cite{amanatidis2024pushing} showed that \(\frac{2}{3}\)-EFX allocations exist when there are at most seven agents or when agents have at most three distinct values for the goods.
\section{Preliminaries}\label{sec:prelims}

Let  \(\N = \{a_1, a_2, \dots, a_n\}\) be a set of \(n\) agents and \(\M = \{g_1, g_2, \dots, g_m\}\) be a set of \(m\) indivisible goods. A discrete fair division instance is represented by the tuple \(\langle \N, \M, \V \rangle\), where \(\V = \langle v_1, v_2, \dots, v_n\rangle\) are the valuation functions. For each agent \(a_i \in \N\), the function \(v_i : 2^{\M} \to \R\) specifies the utility that agent \(a_i\) derives from any given subset of goods.

\begin{definition}[Bundle, Allocation, and Sub-allocation]
We use the term \emph{bundle} to denote a subset of goods. An \emph{allocation} 
\( X = \langle X_{a_1}, X_{a_2}, \dots, X_{a_n} \rangle, \)
over the set of agents \(\N\) is a tuple of \(n\) mutually disjoint bundles, where the bundle \(X_{a_i}\) is assigned to agent \(a_i\) for every \(a_i \in \N\). We say that an allocation \(X\) is \emph{complete} if
\( \bigcup_{a_i \in \N} X_{a_i} = \M \); otherwise it is \emph{partial}. 

Given any allocation \(X\) and any subset \(P \subseteq \N\) of agents, the \emph{sub-allocation} \(X(P)\) is defined as
\( X(P) = \langle X_p \mid p \in P \rangle\), which represents the allocation \(X\) restricted to the agents in \(P\).
\end{definition}

Let \(a\in \N;\  g\in \M;\ S,T\subseteq \M\) and \(v_{a}\) be the valuation function of \(a\). To simplify notation, we write \(v_{a}(g)\) to denote \(v_{a}(\{g\})\) and use \(S\setminus g\), \(S\cup g\) to denote \(S\setminus\{g\}\), \(S\cup\{g\}\), respectively. We also write \(S>_{a} T\) to denote \(v_{a}(S)>v_{a}(T)\) and similarly for \(<_{a}, \geq_{a}, \leq_{a}\) and \(=_{a}\). We use \(\min_{a}(S,T)\) and \(\max_{a}(S,T)\) to denote \(\argmin_{Y\in\{S,T\}} v_{a}(Y)\) and \(\argmax_{Y\in\{S,T\}} v_{a}(Y)\) respectively.

\begin{definition}[Pareto domination]
    An allocation \(Y=\langle Y_{a_1},\ldots,Y_{a_n}\rangle\) \emph{Pareto dominates} an allocation \(X=\langle X_{a_1},\ldots, X_{a_n}\rangle \), denoted by \(Y\succ X\), if for every agent \(a_i\in \N\) her bundle \(Y_{a_i}\) in \(Y\) is at least as good as her bundle \(X_{a_i}\) in \(X\), and for at least one agent \(a_j\), the bundle \(Y_{a_j}\) is strictly better than \(X_{a_j}\). Formally, \(Y\succ X \) if
    \[\forall a_i\in \N, Y_{a_i}\ge_{a_i} X_{a_i} \quad\text{and}\quad \exists a_j\in \N, Y_{a_j}>_{a_j} X_{a_j}.\] 

    Similarly, given two sub-allocations \(Y(P)\) and \(X(P)\) over a subset \(P\subseteq \N\) of agents, we say \(Y(P)\succ X(P)\) if 
    \(\forall a_i\in P, Y_{a_i}\ge_{a_i} X_{a_i} \quad\text{and}\quad \exists a_j\in P, Y_{a_j}>_{a_j} X_{a_j}.\)
\end{definition}

Given an allocation \(X = \langle X_{a_1}, X_{a_2}, \dots, X_{a_n} \rangle\), we say that an agent \(a_i\) \emph{envies} another agent \(a_j\) if \(X_{a_j} >_{a_i} X_{a_i}\). We often also say that the agent \(a_i\) \emph{envies the bundle} \(X_{a_j}\).  For an allocation \(X\), the envy relation among the agents is represented graphically as follows. 
\begin{definition}[Envy Graph]
    Given an allocation \(X\), let \(E_X=(\N,E)\) be a directed graph where the set of vertices of $E_X$ is the set of agents \(\N\), and for every pair \((a_i,a_j)\) of agents, \((a_i,a_j)\in E\) iff agent \(a_i\) envies agent \(a_j\) under the allocation \(X\).
\end{definition}

The envy of an agent could either be strong or weak, as defined below:

\begin{definition}[Strong envy]
Given an allocation \(X\), an agent \(a_i\) \emph{strongly envies} an agent \(a_j\) if \(X_{a_i} <_{a_i} X_{a_j}\setminus g\) for some \(g\in X_{a_j}\).    
\end{definition}

An envy that is not strong is called a \emph{weak} envy. Given the definitions of strong and weak envies, an EFX allocation can be defined as follows.
\begin{definition}[EFX]
    An allocation \(X = \langle X_{a_1}, X_{a_2}, \dots, X_{a_n} \rangle\) is said to be \emph{envy-free up to any good (EFX)} if no agent strongly envies any other agent. That is, \[\forall a_i,a_j\in \N, \forall g\in X_{a_j},\quad X_{a_i} \ge_{a_i} X_{a_j}\setminus g.\]
\end{definition}

 \paragraph{Non-degenerate instances \cite{efx_3,efx_3_simple}} An instance \( \mathcal{I} =\langle \N,\M,\V \rangle\) is said to be \emph{non-degenerate} if no agent values two different bundles equally. That is, \(\forall{a_i \in \N},\ \forall S,T\subseteq \M\), we have \(S\neq T \Rightarrow v_{i}(S) \ne v_{i}(T)\). \cite{efx_3_simple} showed that it suffices to deal with non-degenerate instances when there are \(n\) agents with general valuation functions, i.e., if each non-degenerate instance has an EFX allocation, each general instance has an EFX allocation. In the rest of the paper, we only consider non-degenerate instances. This implies that all goods are positively valued by all agents as the value of the empty bundle is assumed to be zero. 
Non-degenerate instances have the following property.
\begin{proposition}\label{prop:identical_envy}
Given any allocation \(X\), for any two agents \(a\) and \(a'\) with identical valuation, if at least one of them is allocated a non-empty bundle, then either \(a\) envies \(a'\) or \(a'\) envies \(a\).
\end{proposition}

%If an agent \(a_i\) envies a bundle \(S\), we can define a \emph{minimally envied subset} as follows. 
For an envied subset of items, we define a \emph{minimally envied subset} as follows:
\begin{definition}[Minimally envied subset \cite{efx_3}] Given an allocation \(X\) and some \(S \subseteq \M\) that an agent \(a_i\) envies,  %we define a minimum cardinality subset 
a \emph{minimally envied subset} of \(S\) for agent \(a_i\) is defined to be a minimum cardinality subset $T$ of $S$ that $a_i$ envies. %\(T \subseteq S\) as  if: 
Thus, \((i)\  X_{a_i} <_{a_i} T\) and \((ii)\ X_{a_i} \ge_{a_i} T\setminus{h}\quad\forall{h\in T}\). 
\end{definition}
%\begin{definition}[Minimally envied subset \cite{chaudhury2021little}] Given an allocation \(X\) and a bundle \(S\subseteq \M\), if an agent \(a_i\) envies \(S\), we define a subset \(T\subseteq S\) as a \emph{minimally envied subset} of \(S\) for agent \(a_i\) if:  \((i)\) \(X_{a_i} <_{a_i} T\), and \((ii)\) \(X_{a_i} \ge_{a_i} T\setminus{h}\quad\forall{h\in T}\).
%\end{definition}

We also define a minimally envied subset of $S$ for an agent $a_i$ with respect to an arbitrary subset of items $R$ analogously. Thus, for an agent \(a_i\) and \(R, S \subseteq \M\) such that \(R <_{a_i} S\), the set \(T\) is said to be a minimally envied subset of \(S\) with respect to \(R\), denoted \(T = \mes_{a_i}(S, R)\), if \(T\) is the subset of \(S\) that would be minimally envied by \(a_i\) if $R$ is allocated to $a_i$. When \(R\) is the same as \(X_{a_i}\), we simply refer to \(\mes_{a_i}(S, R)\) as \(\mes_{a_i}(S)\).

\begin{definition}[Champion \cite{efx_3}]
   Let \(X\) be an allocation and \(S\subseteq \M\) be envied by at least one agent. An agent \(a_i\) is said to \emph{champion} \(S\) if $a_i=\argmin_{a\in \N} |\mes_a(S)|$ i.e. among all the agents that envy $S$, $a_i$ is an agent with a minimum cardinality minimally envied subset of $S$.%no agent strongly envies the minimally envied subset \(\mes_{a_i}(S)\).
\end{definition} 

\begin{definition}[\(g\)-champion \cite{efx_3}]
    Given a partial allocation \(X\) and an unallocated good \(g\), an agent \(a_i\) is said to \(g\)-champion agent \(a_j\) if \(a_i\) champions the bundle \(X_{a_j}\cup g\). If \(a_i = a_j\), we say agent \(a_i\) self \(g\)-champions. 
\end{definition}

Similar to the envy graph \(E_X\), we can define a \emph{champion graph} with respect to an unallocated good as follows.

\begin{definition}[Champion Graph]\label{def:champion_graph_M_X}
    Given a partial allocation \(X\) along with an unallocated good \(g\), a \emph{champion graph}, denoted by \(M_X^g=(\N,E_M)\), is a directed graph with the vertex set same as the set of agents \(\N\), %as the vertices 
    and for every pair of agents \(a_i,a_j\), the edge \((a_i,a_j)\in E_M\) if \(a_i\) \(g\)-champions \(a_j\).
\end{definition}

The following is a simple observation about champion graphs. 
\begin{proposition}\label{prop:M_X_is_cyclic_basic}
    Let $X$ be a partial allocation and $g$ be an unallocated good. Then, the champion graph $M^g_X$ is always cyclic.
\end{proposition}
\begin{proof}
    Recall that every agent values every good non-zero. For each agent \(a_i\), as \(a_i\) envies \(X_{a_i} \cup g\) over \(X_{a_i}\), some agent (possibly \(a_i\)) must \(g\)-champion \(a_i\). Thus, every agent has an incoming \(g\)-champion edge, making \(M^g_X\) cyclic.
\end{proof}
For simplicity, we refer to \(M_X^g\) as \(M_X\) with the implicit understanding that all champion edges correspond to some unallocated good \(g\), unless explicitly stated otherwise.

\paragraph{Pareto Improvements:} Given an EFX allocation \(X\), which is not necessarily complete, several techniques can be applied to find new EFX allocations that Pareto dominate \(X\). We will discuss some of these techniques below.
Firstly, we can eliminate envy cycles by cyclic exchanges of bundles. This leads to the following lemma: 
\begin{lemma}[\cite{lipton}]
    If \(X\) is an EFX allocation such that the envy graph \(E_X\) of \(X\) has a directed cycle, then there exists another EFX allocation \(Y\) such that \(Y\succ X\).
\end{lemma}

\cite{chaudhury2021little} generalized this idea as follows.
\begin{lemma}[\cite{chaudhury2021little}]\label{lemma:kavita}
    Let \(X\) be a partial EFX allocation. If there exist two agents \(a_i\) and \(a_j\) such that \(a_i\) \(g\)-champions \(a_j\), and \(a_i\) is reachable from \(a_j\) in the envy graph \(E_X\), then there exists another EFX allocation \(Y\) such that \(Y\succ X\). Moreover, all the agents in the path from \(a_i\) to \(a_j\), including \(a_i\) and \(a_j\), get strictly better bundles in \(Y\). 
\end{lemma}
As a corollary of the above lemma, we can make the following two observations.
\begin{corollary}[\cite{chaudhury2021little}]\label{corr:self_g_champion}
    Let \(X\) be a partial EFX allocation. If at least one of the following holds, then there exists another EFX allocation \(Y\) such that \(Y\succ X\). 
    \begin{enumerate}
        \item Envy graph \(E_X\) has a single source vertex.
        \item There exists an agent \(a_i\) who self \(g\)-champions.
    \end{enumerate}
\end{corollary}

Given an allocation \(X\), if agent \(a_i\) does not envy agent \(a_j\) but \(a_i\) \(g\)-champions \(a_j\), it follows that \(g \in \mes_{a_i}(X_{a_j} \cup g, X_{a_i})\). In this case, we say that agent \(a_i\) \emph{decomposes} the bundle \(X_{a_j}\) into two parts, referred to as the \emph{top} and \emph{bottom} halves. The top half is defined as \(T_{a_j}^{a_i} = \mes_{a_i}(X_{a_j} \cup g, X_{a_i}) \setminus g\), while the bottom half is given by \(B_{a_j}^{a_i} = X_{a_j} \setminus T_{a_j}^{a_i}\). This decomposition allows us to compare the top halves of different bundles, leading to the following observation.

\begin{proposition}\label{prop:i_like_the_top_that_i_cut}
    Consider an allocation \(X\). Suppose agent $a_i$ $g$-champions agent $a_j$ and does not \(g\)-champion \(a_\ell\). Suppose agent $a_k$, for $k \neq i$, $g$-champions agent $a_\ell$. The bundle \(X_{a_j}\) is decomposed as \(T_{a_j}^{a_i}\) and \(B_{a_j}^{a_i}\). Similarly, \(X_{a_\ell}\) is decomposed into \(T_{a_\ell}^{a_k}\) and \(B_{a_\ell}^{a_k}\). Then \(T_{a_j}^{a_i} >_{a_i} T_{a_\ell}^{a_k}\).
\end{proposition}
\begin{proof}
    Agent \(a_i\) \(g\)-champions \(a_j\) but not \(a_\ell\). So, \(T_{a_j}^{a_i} \cup g >_{a_i} X_{a_i} >_{a_i} T_{a_\ell}^{a_k} \cup g\), and \(T_{a_j}^{a_i} >_{a_i} T_{a_\ell}^{a_k}\).
\end{proof}

\section{Three Types of Agents}\label{sec:three-types}

Given an instance \(\langle \N, \M, \V \rangle\) of discrete fair division, we say that the agents are of three \emph{types} if the set of agents can be partitioned as \(\N = \A \uplus \B \uplus \C\), where agents in each of the sets \(\A\), \(\B\), and \(\C\) have identical valuation functions. Let \(p,q\) and \(r\) be the number of agents in parts \(\A,\B\) and \(\C\) respectively. Thus, \(|\N|=n=p+q+r\). We refer to these as the three \emph{groups} of agents. In particular, we denote the agents in group \(\A\) as \(\A = \{a_1, a_2, a_3, \dots, a_p\}\), all of whom have the same additive valuation function \(v_a\). Similarly, the agents in group \(\B = \{b_1, b_2, \dots, b_q\}\) and group \(\C = \{c_1, c_2, \dots, c_r\}\) have identical additive valuation functions \(v_b\) and \(v_c\), respectively.

For concise notation, we use \(S>_a T\) as a shorthand for \(S>_{a_i} T\) for agents \(a_i \in \A\). Similarly, we use \(\ge_a, <_a,\le_a,=_a\) and \(\mes_a(\cdot)\) to denote \(\ge_{a_i}, <_{a_i}, \le_{a_i}, =_{a_i}\) and \(\mes_{a_i}(\cdot)\) respectively. Likewise, we use subscripts \(b\) and \(c\) for agents in groups \(\B\) and \(\C\), respectively.

Consider an allocation \(X\) for a non-degenerate instance with three types of agents. For any two agents \(a_i\) and \(a_j\) of the same type, say $\A$, we know that either \(X_{a_i}>_aX_{a_j}\) or \(X_{a_i}<_a X_{a_j}\). Therefore, w.l.o.g.\ we assume that the bundles are sorted among the agents in each group and therefore the agents \(a_1,b_1\) and \(c_1\) have the least valued bundles in their respective groups. We call them the \emph{leading agents}, denoted by \(L=\{a_1,b_1,c_1\}\). Throughout this paper, we will maintain this property as an invariant for each EFX allocation.

\begin{definition}[Ordering Invariant ]\label{def:order-inv} 
    Let $X$ be an EFX allocation. For any group $\mathbb{T}\in \{\A,\B,\C\}$ with valuation $v_t$, if $\exists i>j$ such that $v_t(X_t^i)<v_t(X_t^j)$ then we interchange the bundles of the two agents. Thus, we maintain the invariant that $v_t(X_t^i)\geq v_t(X_t^j)$ whenever $i>j$, for each group $\mathbb{T}$ with valuation function $v_t$. In particular, the leading agent within each group has the minimum-valued bundle according to the valuation function of that group.
\end{definition}

%\subsection{Structure of the Envy and Champion Graphs}
For an instance with three types of agents, we make the following observations on the structure of the envy and champion graphs.
\begin{proposition}\label{prop:M_X_is_cyclic}
     Given an instance with $3$ types of agents, and a partial allocation \(X\), only leading agents can be the sources in the envy graph \(E_X\). Hence $E_X$ has at most $3$ sources. Moreover, the following hold:
     \begin{enumerate}
         \item\label{itm:ordering-envy} If any non-leading agent $x$ envies (respectively $g$-champions) another agent $y$, then the leading agent in the group of $x$ also envies (respectively $g$-champions) $y$. 
         \item\label{itm:ordering-champion} In the champion graph $M_X$ w.r.t.\ an unallocated good $g$, there must be a cycle of length at most $3$, and the cycle involves only the leading agents. 
         % \nv{"must be a cycle of length at most $3$ that involves the leading agents" -- isn't this more accurate?}\pn{ Addressed.}
    \end{enumerate}
 \end{proposition}
 \begin{proof}
    By the Ordering Invariant (Definition~\ref{def:order-inv}), if $x$ envies $y$, the leading agent in the same group as $x$ has the same valuation function and a smaller valued bundle than $x$. Hence Part~\ref{itm:ordering-envy} follows.

    To see Part~\ref{itm:ordering-champion}, consider allocations of $g$ to each of the leading agents. From Proposition~\ref{prop:M_X_is_cyclic_basic}, we know that \(M_X\) is cyclic. If a leading agent self-$g$-champions, then $M_X$ has a cycle of length $1$. If no leading agent self-$g$-champions, then by Part~\ref{itm:ordering-envy}, no agent in any group $g$-champions the leading agent in the same group. So the leading agent of each group must be $g$-championed by a leading agent of another group, thereby creating a $2$-cycle or a $3$-cycle.
\end{proof}

\subsection{Progress Measures}\label{subsec:progress}

A common approach to proving the existence of EFX allocations is to begin with an empty, trivially EFX allocation, and iteratively {\em improve} it. To quantify the improvement and ensure termination, various \emph{progress measures} have been proposed. Pareto dominance is one such well-studied measure, where each partial EFX allocation Pareto dominates the previous one. However, \cite{efx_3} showed that this approach fails to guarantee EFX even for three agents. Instead, they used \emph{lexicographic} improvement to prove the existence of EFX for three agents. 

To prove Theorem~\ref{thm:efx3}, we use a combination of Pareto dominance and a new potential function defined below. 

% \textcolor{blue}{Look at this def again.}\textcolor{red}{PN: This looks fine to me.}

\begin{definition}[Potential Function]\label{def:potential} For any (possibly partial) allocation \(X\), the potential of \(X\) is defined 
%with respect to a fixed group, say \(\A\), 
as:  \[ \phi(X) = \min_{a_i \in \A} v_a(X_{a_i}). \]  
\end{definition}

{We mention that the choice of group $\A$ in the definition of $\phi$ is arbitrary. However, the structure of the proof of the main result depends on the identity of the group of agents with respect to which the potential function is defined, which, in this case is $\A$.}

Note that the function \(\phi\) is \emph{compatible} with Pareto dominance. That is, 
\begin{proposition}\label{prop:pareto_compatible_phi}
    For allocations \(X\), \(Y\), if \(Y\succ X\), then \(\phi(Y) \ge \phi(X)\).
\end{proposition}

If \(\phi(Y)>\phi(X)\), we say \(Y\) \(\phi\)-dominates \(X\). We start with a trivial EFX allocation and iteratively find new EFX allocations that either Pareto dominate or \(\phi\)-dominate the preceding one. As Pareto dominance is a partial order on the set of allocations, and \(\phi\) forms a total order, this procedure must terminate. 

When we encounter allocations such that the strong envies are confined to agents within the same group(s), that is, there is no \emph{inter-group} strong envy, such envy can be resolved without reducing the potential \(\phi\). This is shown more generally in the following lemma.

%\textcolor{blue}{Shouldn't we also say that no change happens to the allocation of agents in the other groups? }

\begin{restatable}[Resolving internal strong envies]{lemma}{internalEnvy}\label{lemma:internal_envies}
    Let \(X\) be any allocation. Suppose an arbitrary agent, say
    %\(a_i\in \A\), 
    $t \in \mathbb{T}$ for $\mathbb{T} \in \{\A,\B,\C\}$,
    strongly envies a set $S$ of agents only within her group, i.e., \(S \subset \mathbb{T}\). Then, there exists another allocation \(Y\) such that the following conditions hold:
  \begin{enumerate}
      \item \(\phi(Y) = \phi(X)\)
      \item In \(Y\), agent \(t\) does not strongly envy anyone.
      \item In \(Y\), no agent in \(S\) strongly envies any agent.
      \item Bundles of agents in the other two groups remain unchanged.
  \end{enumerate}
\end{restatable}
\begin{proof} 
Without loss of generality, let $\mathbb{T} = \A$ and $t = a_i$.
    Let $a_j\in S$. So $a_i$ strongly envies $a_j$. Then we replace $X_{a_j}$ by $Y_{a_j}=\mes_{a_i}(X_{a_j})$. Clearly $X_{a_j}$ is not the minimum valued bundle in $\A$. So $\phi$ doesn't change by this replacement. Since such a replacement is done for each $a_j\in S$, $a_i$ has no strong envy towards any agent in $S$. Since $a_i$ did not strongly envy anyone in other groups with respect to $X$, and since all the agents outside $S$ have the same bundles in $X$ and $Y$, $a_i$ does not strongly envy anyone outside $S$ either. Although agents in $S$, like $a_j$, got smaller bundles, $Y_{a_j}>_a X_{a_i}$. So $a_j$ does not strongly envy anyone with respect to $Y$ as $a_i$ does not strongly envy anyone with respect to $Y$.\qedhere
\end{proof}

Since Lemma~\ref{lemma:internal_envies} applies to any of the three groups, we have the following corollary.
\begin{corollary}\label{corr:intra-group}
    If \(X\) is an allocation in which no agent strongly envies another agent from a different group,  there exists an EFX allocation \(Y\), possibly partial, such that \(\phi(Y) = \phi(X)\).
\end{corollary}

\subsection{Pseudo-Cycles}\label{subsec:pseudo_cycles}
%We generalize this idea to cases where there is no actual cycle among the agents, but rather among the groups. We call this a \emph{pseudo-cycle}, as discussed in the following section. 

Similar to Pareto Improving Cycles in \cite{BergerCFF22}, we define pseudo-cycles for instances with three types of agents. These are not necessarily cycles among the agents, but rather cycles among the groups. Some examples of pseudo-cycles are given in Figure~\ref{fig:phi_cycles}.

\begin{definition}[Pseudo-Cycle]\label{def:pseudo-cycle}
    Given a possibly partial EFX allocation \(X\), a pseudo-cycle involving three groups is a tuple of edges either of the form \(\langle (a_1,b_i),(b_1,c_j),(c_1,a_1)\rangle\) or of the form \(\langle(a_1,c_j),(c_1,b_i),(b_1,a_1)\rangle\), where \(b_i\in\B\) and \(c_j\in \C\), such that the following holds:
    \begin{enumerate}
        \item every edge in the tuple is either an envy edge or a \(g\)-champion edge for some unallocated good \(g\).
        \item For each unallocated good \(g\), the tuple can have at most one \(g\)-champion edge.
    \end{enumerate}

    Similarly, a pseudo-cycle involving two groups is a tuple of the form \(\langle(a_1,b_i),(b_1,a_1)\rangle\) (or \(\langle(a_1,c_j),(c_1,a_1)\rangle\)) such that the above conditions hold. 
\end{definition}

\begin{figure}[ht]
    \centering
    \begin{subfigure}[b]{0.64\textwidth}
        \centering
        \input{figures/example3}
        \caption{\(\langle(a_1,b_i),(b_1,c_j),(c_1,a_1)\rangle\) is a pseudo-cycle with only envy edges}
        \label{fig:c}
    \end{subfigure}
    \hfill
    \begin{subfigure}[b]{0.34\textwidth}
        \centering
        \tikzset{every picture/.style={line width=0.75pt}} %set default line width to 0.75pt        

\begin{tikzpicture}[x=0.75pt,y=0.75pt,yscale=-1,xscale=1]
%uncomment if require: \path (0,109); %set diagram left start at 0, and has height of 109

%Shape: Ellipse [id:dp30887097265558827] 
\draw  [color={rgb, 255:red, 93; green, 166; blue, 13 }  ,draw opacity=1 ][fill={rgb, 255:red, 232; green, 254; blue, 212 }  ,fill opacity=1 ][line width=0.75]  (399.41,24.54) .. controls (399.48,29.41) and (395.47,33.42) .. (390.46,33.49) .. controls (385.45,33.56) and (381.33,29.67) .. (381.25,24.8) .. controls (381.18,19.93) and (385.19,15.92) .. (390.2,15.85) .. controls (395.21,15.77) and (399.33,19.66) .. (399.41,24.54) -- cycle ;
%Shape: Ellipse [id:dp17509738642631667] 
\draw  [color={rgb, 255:red, 0; green, 0; blue, 0 }  ,draw opacity=1 ][fill={rgb, 255:red, 221; green, 221; blue, 221 }  ,fill opacity=1 ][line width=0.75]  (359.07,25.01) .. controls (359.14,29.83) and (355.19,33.79) .. (350.24,33.86) .. controls (345.29,33.93) and (341.22,30.09) .. (341.15,25.27) .. controls (341.08,20.46) and (345.03,16.5) .. (349.98,16.43) .. controls (354.93,16.36) and (359,20.2) .. (359.07,25.01) -- cycle ;
%Shape: Ellipse [id:dp38763608257879456] 
\draw  [color={rgb, 255:red, 0; green, 0; blue, 0 }  ,draw opacity=1 ][fill={rgb, 255:red, 221; green, 221; blue, 221 }  ,fill opacity=1 ][line width=0.75]  (273.93,26.25) .. controls (274,31.06) and (270.04,35.02) .. (265.09,35.09) .. controls (260.14,35.17) and (256.07,31.32) .. (256,26.51) .. controls (255.93,21.7) and (259.89,17.74) .. (264.84,17.67) .. controls (269.79,17.59) and (273.86,21.44) .. (273.93,26.25) -- cycle ;
%Shape: Ellipse [id:dp8606058797405975] 
\draw  [color={rgb, 255:red, 0; green, 0; blue, 0 }  ,draw opacity=1 ][fill={rgb, 255:red, 221; green, 221; blue, 221 }  ,fill opacity=1 ][line width=0.75]  (318.96,25.49) .. controls (319.03,30.3) and (315.08,34.26) .. (310.13,34.33) .. controls (305.18,34.4) and (301.11,30.56) .. (301.04,25.75) .. controls (300.97,20.93) and (304.92,16.97) .. (309.87,16.9) .. controls (314.82,16.83) and (318.89,20.67) .. (318.96,25.49) -- cycle ;
%Straight Lines [id:da1240974796710469] 
\draw [color={rgb, 255:red, 0; green, 0; blue, 0 }  ,draw opacity=1 ][line width=0.75]    (381.25,24.8) -- (362.07,24.98) ;
\draw [shift={(359.07,25.01)}, rotate = 359.45] [fill={rgb, 255:red, 0; green, 0; blue, 0 }  ,fill opacity=1 ][line width=0.08]  [draw opacity=0] (7.14,-3.43) -- (0,0) -- (7.14,3.43) -- cycle    ;
%Straight Lines [id:da6525973917811535] 
\draw [color={rgb, 255:red, 0; green, 0; blue, 0 }  ,draw opacity=1 ][line width=0.75]    (341.15,25.27) -- (321.96,25.46) ;
\draw [shift={(318.96,25.49)}, rotate = 359.45] [fill={rgb, 255:red, 0; green, 0; blue, 0 }  ,fill opacity=1 ][line width=0.08]  [draw opacity=0] (7.14,-3.43) -- (0,0) -- (7.14,3.43) -- cycle    ;
%Straight Lines [id:da6707404436534767] 
\draw [color={rgb, 255:red, 0; green, 0; blue, 0 }  ,draw opacity=1 ][line width=0.75]    (287.37,26.05) -- (276.93,26.21) ;
\draw [shift={(273.93,26.25)}, rotate = 359.17] [fill={rgb, 255:red, 0; green, 0; blue, 0 }  ,fill opacity=1 ][line width=0.08]  [draw opacity=0] (7.14,-3.43) -- (0,0) -- (7.14,3.43) -- cycle    ;
%Shape: Ellipse [id:dp8458876747999241] 
\draw  [color={rgb, 255:red, 93; green, 166; blue, 13 }  ,draw opacity=1 ][fill={rgb, 255:red, 232; green, 254; blue, 212 }  ,fill opacity=1 ][line width=0.75]  (256.6,79.31) .. controls (256.52,74.43) and (260.53,70.43) .. (265.54,70.35) .. controls (270.55,70.28) and (274.68,74.17) .. (274.75,79.04) .. controls (274.82,83.92) and (270.81,87.93) .. (265.8,88) .. controls (260.79,88.07) and (256.67,84.18) .. (256.6,79.31) -- cycle ;
%Shape: Ellipse [id:dp5695430461065334] 
\draw  [color={rgb, 255:red, 0; green, 0; blue, 0 }  ,draw opacity=1 ][fill={rgb, 255:red, 221; green, 221; blue, 221 }  ,fill opacity=1 ][line width=0.75]  (296.93,78.83) .. controls (296.86,74.02) and (300.81,70.06) .. (305.76,69.99) .. controls (310.71,69.92) and (314.78,73.76) .. (314.86,78.57) .. controls (314.93,83.38) and (310.97,87.34) .. (306.02,87.41) .. controls (301.07,87.49) and (297,83.64) .. (296.93,78.83) -- cycle ;
%Shape: Ellipse [id:dp608752812961753] 
\draw  [color={rgb, 255:red, 0; green, 0; blue, 0 }  ,draw opacity=1 ][fill={rgb, 255:red, 221; green, 221; blue, 221 }  ,fill opacity=1 ][line width=0.75]  (382.07,77.59) .. controls (382,72.78) and (385.96,68.82) .. (390.91,68.75) .. controls (395.86,68.68) and (399.93,72.52) .. (400,77.33) .. controls (400.07,82.15) and (396.11,86.11) .. (391.16,86.18) .. controls (386.21,86.25) and (382.14,82.41) .. (382.07,77.59) -- cycle ;
%Shape: Ellipse [id:dp39439673822440713] 
\draw  [color={rgb, 255:red, 0; green, 0; blue, 0 }  ,draw opacity=1 ][fill={rgb, 255:red, 221; green, 221; blue, 221 }  ,fill opacity=1 ][line width=0.75]  (337.04,78.36) .. controls (336.97,73.54) and (340.92,69.59) .. (345.87,69.51) .. controls (350.82,69.44) and (354.89,73.28) .. (354.96,78.1) .. controls (355.03,82.91) and (351.08,86.87) .. (346.13,86.94) .. controls (341.18,87.01) and (337.11,83.17) .. (337.04,78.36) -- cycle ;
%Straight Lines [id:da2526226729904757] 
\draw [color={rgb, 255:red, 0; green, 0; blue, 0 }  ,draw opacity=1 ][line width=0.75]    (274.74,79.04) -- (293.93,78.86) ;
\draw [shift={(296.93,78.83)}, rotate = 179.45] [fill={rgb, 255:red, 0; green, 0; blue, 0 }  ,fill opacity=1 ][line width=0.08]  [draw opacity=0] (7.14,-3.43) -- (0,0) -- (7.14,3.43) -- cycle    ;
%Straight Lines [id:da21909450316962564] 
\draw [color={rgb, 255:red, 0; green, 0; blue, 0 }  ,draw opacity=1 ][line width=0.75]    (314.85,78.57) -- (334.04,78.39) ;
\draw [shift={(337.04,78.36)}, rotate = 179.45] [fill={rgb, 255:red, 0; green, 0; blue, 0 }  ,fill opacity=1 ][line width=0.08]  [draw opacity=0] (7.14,-3.43) -- (0,0) -- (7.14,3.43) -- cycle    ;
%Straight Lines [id:da5056033919268932] 
\draw [color={rgb, 255:red, 0; green, 0; blue, 0 }  ,draw opacity=1 ][line width=0.75]    (368.63,77.79) -- (379.07,77.64) ;
\draw [shift={(382.07,77.59)}, rotate = 179.17] [fill={rgb, 255:red, 0; green, 0; blue, 0 }  ,fill opacity=1 ][line width=0.08]  [draw opacity=0] (7.14,-3.43) -- (0,0) -- (7.14,3.43) -- cycle    ;
%Curve Lines [id:da8240465293644047] 
\draw [color={rgb, 255:red, 208; green, 2; blue, 27 }  ,draw opacity=1 ]   (390.46,33.49) .. controls (383.26,50.18) and (355.07,44.25) .. (346.7,66.93) ;
\draw [shift={(345.87,69.51)}, rotate = 285.48] [fill={rgb, 255:red, 208; green, 2; blue, 27 }  ,fill opacity=1 ][line width=0.08]  [draw opacity=0] (7.14,-3.43) -- (0,0) -- (7.14,3.43) -- cycle    ;
%Curve Lines [id:da09784556940794165] 
\draw [color={rgb, 255:red, 0; green, 0; blue, 0 }  ,draw opacity=1 ]   (272,72) .. controls (282.2,54.87) and (353.31,51.18) .. (381.89,33.4) ;
\draw [shift={(384,32)}, rotate = 144.87] [fill={rgb, 255:red, 0; green, 0; blue, 0 }  ,fill opacity=1 ][line width=0.08]  [draw opacity=0] (7.14,-3.43) -- (0,0) -- (7.14,3.43) -- cycle    ;

% Text Node
\draw (302.97,29.7) node [anchor=north west][inner sep=0.75pt]  [font=\scriptsize,rotate=-180]  {$\cdots $};
% Text Node
\draw (383.55,20.52) node [anchor=north west][inner sep=0.75pt]  [font=\scriptsize] [align=left] {$\displaystyle a_{1}$};
% Text Node
\draw (351.87,74.14) node [anchor=north west][inner sep=0.75pt]  [font=\scriptsize]  {$\cdots $};
% Text Node
\draw (259.82,73.14) node [anchor=north west][inner sep=0.75pt]  [font=\scriptsize] [align=left] {$\displaystyle b_{1}$};
% Text Node
\draw (339.82,73.14) node [anchor=north west][inner sep=0.75pt]  [font=\scriptsize] [align=left] {$\displaystyle b_{i}$};
\draw (360,50.4) node [anchor=north west][inner sep=0.75pt]  [font=\scriptsize,color={rgb, 255:red, 208; green, 2; blue, 27 }  ,opacity=1 ]  {$g$};

\end{tikzpicture}
        \caption{\(\langle(a_1,b_i),(b_1,a_1)\rangle\) is a pseudo-cycle}
        \label{fig:d}
    \end{subfigure}
    \centering
        \begin{subfigure}[b]{0.55\textwidth}
            \centering
            \input{figures/example1}
            \caption{\(\langle(a_1,b_i),(b_1,c_j),(c_1,a_1)\rangle\) is a pseudo cycle involving two compatible champion edges.}
            \label{fig:a}
        \end{subfigure}
        \hfill
        \begin{subfigure}[b]{0.39\textwidth}
            \centering
            \tikzset{every picture/.style={line width=0.75pt}} %set default line width to 0.75pt        

\begin{tikzpicture}[x=0.75pt,y=0.75pt,yscale=-1,xscale=1]
%uncomment if require: \path (0,114); %set diagram left start at 0, and has height of 114

%Shape: Ellipse [id:dp5864386992981451] 
\draw  [color={rgb, 255:red, 93; green, 166; blue, 13 }  ,draw opacity=1 ][fill={rgb, 255:red, 232; green, 254; blue, 212 }  ,fill opacity=1 ][line width=0.75]  (267.24,41) .. controls (267.31,45.87) and (263.3,49.88) .. (258.29,49.95) .. controls (253.28,50.03) and (249.16,46.13) .. (249.09,41.26) .. controls (249.02,36.39) and (253.02,32.38) .. (258.03,32.31) .. controls (263.05,32.23) and (267.17,36.13) .. (267.24,41) -- cycle ;
%Shape: Ellipse [id:dp5136069383060925] 
\draw  [color={rgb, 255:red, 0; green, 0; blue, 0 }  ,draw opacity=1 ][fill={rgb, 255:red, 221; green, 221; blue, 221 }  ,fill opacity=1 ][line width=0.75]  (221.93,42.16) .. controls (222,46.97) and (218.04,50.93) .. (213.09,51) .. controls (208.14,51.07) and (204.07,47.23) .. (204,42.41) .. controls (203.93,37.6) and (207.88,33.64) .. (212.84,33.57) .. controls (217.79,33.5) and (221.86,37.34) .. (221.93,42.16) -- cycle ;
%Straight Lines [id:da9744309489054815] 
\draw [color={rgb, 255:red, 0; green, 0; blue, 0 }  ,draw opacity=1 ][line width=0.75]    (235.37,41.96) -- (224.93,42.11) ;
\draw [shift={(221.93,42.16)}, rotate = 359.17] [fill={rgb, 255:red, 0; green, 0; blue, 0 }  ,fill opacity=1 ][line width=0.08]  [draw opacity=0] (7.14,-3.43) -- (0,0) -- (7.14,3.43) -- cycle    ;
%Shape: Ellipse [id:dp9879346688563231] 
\draw  [color={rgb, 255:red, 93; green, 166; blue, 13 }  ,draw opacity=1 ][fill={rgb, 255:red, 232; green, 254; blue, 212 }  ,fill opacity=1 ][line width=0.75]  (306.6,41.65) .. controls (306.52,36.78) and (310.53,32.77) .. (315.54,32.7) .. controls (320.55,32.63) and (324.68,36.52) .. (324.75,41.39) .. controls (324.82,46.26) and (320.81,50.27) .. (315.8,50.34) .. controls (310.79,50.42) and (306.67,46.53) .. (306.6,41.65) -- cycle ;
%Shape: Ellipse [id:dp7644675063879327] 
\draw  [color={rgb, 255:red, 0; green, 0; blue, 0 }  ,draw opacity=1 ][fill={rgb, 255:red, 221; green, 221; blue, 221 }  ,fill opacity=1 ][line width=0.75]  (346.93,41.18) .. controls (346.86,36.36) and (350.81,32.4) .. (355.76,32.33) .. controls (360.71,32.26) and (364.78,36.1) .. (364.86,40.92) .. controls (364.93,45.73) and (360.97,49.69) .. (356.02,49.76) .. controls (351.07,49.83) and (347,45.99) .. (346.93,41.18) -- cycle ;
%Shape: Ellipse [id:dp5808814139268504] 
\draw  [color={rgb, 255:red, 0; green, 0; blue, 0 }  ,draw opacity=1 ][fill={rgb, 255:red, 221; green, 221; blue, 221 }  ,fill opacity=1 ][line width=0.75]  (432.07,39.94) .. controls (432,35.13) and (435.96,31.17) .. (440.91,31.1) .. controls (445.86,31.02) and (449.93,34.87) .. (450,39.68) .. controls (450.07,44.49) and (446.11,48.45) .. (441.16,48.52) .. controls (436.21,48.6) and (432.14,44.75) .. (432.07,39.94) -- cycle ;
%Shape: Ellipse [id:dp550399926527097] 
\draw  [color={rgb, 255:red, 0; green, 0; blue, 0 }  ,draw opacity=1 ][fill={rgb, 255:red, 221; green, 221; blue, 221 }  ,fill opacity=1 ][line width=0.75]  (387.04,40.7) .. controls (386.97,35.89) and (390.92,31.93) .. (395.87,31.86) .. controls (400.82,31.79) and (404.89,35.63) .. (404.96,40.44) .. controls (405.03,45.26) and (401.08,49.21) .. (396.13,49.29) .. controls (391.18,49.36) and (387.11,45.52) .. (387.04,40.7) -- cycle ;
%Straight Lines [id:da49781030829826356] 
\draw [color={rgb, 255:red, 0; green, 0; blue, 0 }  ,draw opacity=1 ][line width=0.75]    (324.74,41.39) -- (343.93,41.2) ;
\draw [shift={(346.93,41.18)}, rotate = 179.45] [fill={rgb, 255:red, 0; green, 0; blue, 0 }  ,fill opacity=1 ][line width=0.08]  [draw opacity=0] (7.14,-3.43) -- (0,0) -- (7.14,3.43) -- cycle    ;
%Straight Lines [id:da7359625866698509] 
\draw [color={rgb, 255:red, 0; green, 0; blue, 0 }  ,draw opacity=1 ][line width=0.75]    (364.85,40.92) -- (384.04,40.73) ;
\draw [shift={(387.04,40.7)}, rotate = 179.45] [fill={rgb, 255:red, 0; green, 0; blue, 0 }  ,fill opacity=1 ][line width=0.08]  [draw opacity=0] (7.14,-3.43) -- (0,0) -- (7.14,3.43) -- cycle    ;
%Straight Lines [id:da7660133303287596] 
\draw [color={rgb, 255:red, 0; green, 0; blue, 0 }  ,draw opacity=1 ][line width=0.75]    (418.63,40.14) -- (429.07,39.98) ;
\draw [shift={(432.07,39.94)}, rotate = 179.17] [fill={rgb, 255:red, 0; green, 0; blue, 0 }  ,fill opacity=1 ][line width=0.08]  [draw opacity=0] (7.14,-3.43) -- (0,0) -- (7.14,3.43) -- cycle    ;
%Shape: Ellipse [id:dp4254070786608477] 
\draw  [color={rgb, 255:red, 93; green, 166; blue, 13 }  ,draw opacity=1 ][fill={rgb, 255:red, 232; green, 254; blue, 212 }  ,fill opacity=1 ][line width=0.75]  (281.86,84.31) .. controls (281.79,79.43) and (285.79,75.43) .. (290.81,75.35) .. controls (295.82,75.28) and (299.94,79.17) .. (300.01,84.04) .. controls (300.08,88.92) and (296.08,92.93) .. (291.07,93) .. controls (286.05,93.07) and (281.93,89.18) .. (281.86,84.31) -- cycle ;
%Shape: Ellipse [id:dp9393668071995426] 
\draw  [color={rgb, 255:red, 0; green, 0; blue, 0 }  ,draw opacity=1 ][fill={rgb, 255:red, 221; green, 221; blue, 221 }  ,fill opacity=1 ][line width=0.75]  (328.34,82.41) .. controls (328.26,77.6) and (332.22,73.64) .. (337.17,73.57) .. controls (342.12,73.5) and (346.19,77.34) .. (346.26,82.15) .. controls (346.33,86.97) and (342.38,90.93) .. (337.43,91) .. controls (332.48,91.07) and (328.41,87.23) .. (328.34,82.41) -- cycle ;
%Straight Lines [id:da27911237867441674] 
\draw [color={rgb, 255:red, 0; green, 0; blue, 0 }  ,draw opacity=1 ][line width=0.75]    (314.89,84.03) -- (325.34,83.88) ;
\draw [shift={(328.34,83.84)}, rotate = 179.17] [fill={rgb, 255:red, 0; green, 0; blue, 0 }  ,fill opacity=1 ][line width=0.08]  [draw opacity=0] (7.14,-3.43) -- (0,0) -- (7.14,3.43) -- cycle    ;
%Curve Lines [id:da7623533035595415] 
\draw    (258.03,32.31) .. controls (278.9,7.19) and (364.82,7.1) .. (389.41,29) ;
\draw [shift={(391.52,31.09)}, rotate = 228.35] [fill={rgb, 255:red, 0; green, 0; blue, 0 }  ][line width=0.08]  [draw opacity=0] (7.14,-3.43) -- (0,0) -- (7.14,3.43) -- cycle    ;
%Curve Lines [id:da9656997829635505] 
\draw [color={rgb, 255:red, 208; green, 2; blue, 27 }  ,draw opacity=1 ]   (281.86,84.31) .. controls (268.32,87.87) and (255.74,79.61) .. (257.99,52.93) ;
\draw [shift={(258.29,49.95)}, rotate = 96.79] [fill={rgb, 255:red, 208; green, 2; blue, 27 }  ,fill opacity=1 ][line width=0.08]  [draw opacity=0] (7.14,-3.43) -- (0,0) -- (7.14,3.43) -- cycle    ;
%Curve Lines [id:da9285326468572523] 
\draw    (306.6,41.65) .. controls (291.7,43.87) and (292.64,55.92) .. (291.1,72.43) ;
\draw [shift={(290.81,75.35)}, rotate = 276.34] [fill={rgb, 255:red, 0; green, 0; blue, 0 }  ][line width=0.08]  [draw opacity=0] (7.14,-3.43) -- (0,0) -- (7.14,3.43) -- cycle    ;

% Text Node
\draw (250.96,45.8) node [anchor=north west][inner sep=0.75pt]  [font=\scriptsize,rotate=-180]  {$\cdots $};
% Text Node
\draw (251.88,36.98) node [anchor=north west][inner sep=0.75pt]  [font=\scriptsize] [align=left] {$\displaystyle a_{1}$};
% Text Node
\draw (402.87,36.39) node [anchor=north west][inner sep=0.75pt]  [font=\scriptsize]  {$\cdots $};
% Text Node
\draw (309.82,34.89) node [anchor=north west][inner sep=0.75pt]  [font=\scriptsize] [align=left] {$\displaystyle b_{1}$};
% Text Node
\draw (389.82,34.89) node [anchor=north west][inner sep=0.75pt]  [font=\scriptsize] [align=left] {$\displaystyle b_{i}$};
% Text Node
\draw (298.87,80.6) node [anchor=north west][inner sep=0.75pt]  [font=\scriptsize]  {$\cdots $};
% Text Node
\draw (284.88,79.28) node [anchor=north west][inner sep=0.75pt]  [font=\scriptsize] [align=left] {$\displaystyle c_{1}$};
% Text Node
\draw (250,70.4) node [anchor=north west][inner sep=0.75pt]  [font=\scriptsize,color={rgb, 255:red, 208; green, 2; blue, 27 }  ,opacity=1 ]  {$g$};

\end{tikzpicture}
            \caption{\(\langle(a_1,b_i),(b_1,c_1),(c_1,a_1)\rangle\) is a pseudo cycle involving one \(g\)-champion edge}
            \label{fig:b}
        \end{subfigure}
        
    \caption{Some examples of pseudo-cycles}
    \label{fig:phi_cycles}
\end{figure}

\subsection{Elementary Cases for Pareto and \(\phi\)-Improvements}\label{sec:simple}
Given a partial EFX allocation \(X\), we discuss a few simple methods to obtain a Pareto dominating or a  \(\phi\)-dominating EFX allocation. 

{\em Elimination of a pseudo-cycle:} A pseudo-cycle is eliminated by giving the envied bundle to the envying group. Thus a group that loses a bundle also receives a bundle. While the received bundle may have a lower value than the lost bundle, note that it is still higher valued than the minimum bundle in that group. Within the group, the newly received bundle is then allocated to an agent, with possible shuffling of existing bundles, so as to maintain the ordering invariant. As $a_1$ is always an envying and envied agent in a pseudo-cycle, the following lemma holds.

\begin{restatable}{lemma}{phiExchange}\label{lemma:pseudo-cycle}
    Eliminating a pseudo-cycle in an EFX allocation \(X\) gives another EFX allocation \(Y\) such that \(\phi(Y)>\phi(X)\).
\end{restatable}
\begin{proof}
    Consider an EFX allocation \(X\). A pseudo-cycle in \(X\) can either involve agents from two groups, or from all the three groups. 

    If it involves two groups, say \(\langle (a_1,b_i),(b_1,a_1) \rangle\), we form a new allocation \(Y\) by assigning \(X_{a_1}\) to \(b_1\) (or \(\mes_b(X_{a_1} \cup g)\) if \((b_1,a_1)\) is a \(g\)-champion edge) and \(X_{b_i}\) (or \(\mes_a(X_{b_i} \cup h)\) for an \(h\)-champion edge) to \(a_i\).

    If it involves three groups, w.l.o.g let \(\langle (a_1,b_i), (b_1,c_j), (c_1,a_1) \rangle\) be a pseudo-cycle. The new allocation \(Y\) is obtained by transferring the bundle of \(b_i\) to \(a_1\), that of \(c_j\) to \(b_1\), and the bundle of \(a_1\) to \(c_1\). If any of these are champion edges, we allocate \(\mes\) bundles accordingly.
    
    The bundles are then re-sorted to restore the Ordering Invariant. Let $X(\A)$ be the set of bundles of agents in $\A$ in the allocation $X$. 
    In allocation $Y$, the value of the minimum valued bundle in $\A$\ increases since the new bundles in $\A$ i.e. $Y(\A)=(X(\A)\setminus \{X_{a_1})\}\cup \{X_{b_i}\}$, and $v_a(X_{b_i})>v_a(X_{a_1})$. So the new minimum valued bundle in $\A$ is $\min\{X_{a_2}, X_{b_i}\}$. 

    Now we prove that $Y$ is an EFX allocation. For each leading agent in \(Y\), the bundle in $Y$ is either the same as that in $X$, or better than $X$. Also, the exchanged bundles are either the same bundles as in $X$ or are an \(\mes\) bundle in $X$. Therefore, the exchanged bundles are not strongly envied by any of the leading agents. Furthermore, the agents whose utility decreased in \(Y\)  still have a higher utility compared to their respective leading agents. Therefore, by applying Proposition~\ref{prop:M_X_is_cyclic}, none of the non-leading agents have strong envy towards any agent. This proves that $Y$ is also an EFX allocation.  \qedhere
\end{proof}

Another way to find a \(\phi\)-dominating EFX allocation is when agent \(a_1\) is the only source in the envy graph \(E_X\). In this case, as a corollary of Lemma~\ref{lemma:kavita}, we obtain a \(\phi\)-dominating EFX allocation. This holds because the bundle value of agent \(a_1\) strictly increases.

\begin{corollary}\label{corr:single_source_phi_improvement}
    Given a partial EFX allocation \(X\), if \(a_1\) is the only source in \(E_X\), then there exists an EFX allocation \(Y\), possibly partial, such that \(\phi(Y)>\phi(X)\).
\end{corollary}

There are cases where two of the leading agents champion each other, while a third agent envies neither of them; in these cases, we can make either Pareto improvements or \(\phi\)-improvements, using the following lemma: 
\begin{restatable}{lemma}{mxTwoCycle}
\label{lemma:mx_two_cycle}
    Let \(X\) be a partial EFX allocation with an unallocated good \(g\) such that in \(X\), two of the leading agents (say, \(b_1\) and \(c_1\)) \(g\)-champion each other and none of the leading agents envy those two agents ($b_1$ and $c_1$, in this case). Then, there exists an EFX allocation \(Y\) such that \(Y \succ X\). Moreover, the bundles of both the leading agents in the \(g\)-champion cycle are strictly improved.
\end{restatable}
\begin{proof}
    Without loss of generality, assume that agents \(b_1\) and \(c_1\) \(g\)-champion each other. Then, the bundles of \(b_1\) and \(c_1\) can be partitioned into top and bottom parts as follows:
    \[ X \ =\  \raisebox{-0.9em}{\hbox{\(\Allocationfiveone{X_{a_1}}{T_{b_1}^{c_1}}{B_{b_1}^{c_1}}{T_{c_1}^{b_1}}{B_{c_1}^{b_1}}\)}} \]
    We consider a new allocation by exchanging the two top halves as follows:
    \[ X'\ =\  \raisebox{-0.9em}{\hbox{\(\Allocationfiveone{X_{a_1}}{T_{c_1}^{b_1}}{B_{b_1}^{c_1}}{T_{b_1}^{c_1}}{B_{c_1}^{b_1}}\)}} \]
    
    From Proposition~\ref{prop:i_like_the_top_that_i_cut}, we know that \(T_{b_1}^{c_1} >_{c} T_{c_1}^{b_1}\) and \(T_{c_1}^{b_1} >_{b} T_{b_1}^{c_1} \). Therefore, both agents \(b_1\) and \(c_1\) are strictly better off. Now, consider the potential strong envies in \(X'\). Since the allocation \(X\) was EFX, any strong envies in \(X'\) must be towards \(b_1\) or \(c_1\). As \(T_{b_1}^{c_1} >_{c} T_{c_1}^{b_1}\), we know that 

        \[X'_{c_1} =_c T_{b_1}^{c_1}\cup B_{c_1}^{b_1} >_c T_{c_1}^{b_1}\cup B_{c_1}^{b_1} =_c X_{c_1}>_c T_{b_1}^{c_1}\cup B_{b_1}^{c_1} =_c X_{b_1}>_c T_{c_1}^{b_1}\cup B_{b_1}^{c_1} =_c X'_{b_1}\]

    Therefore, no agent from group \(\C\) envies agent \(b_1\). Similarly, no one from \(\B\) envies agent \(c_1\). So, the only possible strong envies are from agents in \(\A\) to either \(b_1\) or \(c_1\). We know that if any agent in \(\A\) envies some agent \(a^*\in \N\), then \(a_1\) also envies \(a^*\). Since agent \(a_1\) envied neither \(b_1\) nor \(c_1\) in the allocation \(X\), agent \(a_1\) can envy at most one agent among \(b_1\) and \(c_1\) in \(X'\). If \(a_1\) envies neither, then we arbitrarily pick \(b_1\). Otherwise, we pick the agent that \(a_1\) envies, say \(b_1\).We replace \(B_{b_1}^{c_1}\) in agent \(b_1\)'s bundle with \(g\) to get the following allocation:
    \[ Y \ =\  \raisebox{-0.9em}{\hbox{\(\Allocationfiveone{X_1}{T_{c_1}^{b_1}}{g}{T_{b_1}^{c_1}}{B_{c_1}^{b_1}}\)}} \]
    
    As \(T_{c_1}^{b_1}\cup g\) is a minimally envied subset, no agent strongly envies \(b_1\). Since \(T_{c_1}^{b_1}\cup g >_{b} X_{b_1}\), she does not envy agent \(c_1\). Furthermore, since agent \(c_1\) did not self \(g\)-champion in \(X\), no agent in \(\C\) envy agent \(b_1\). Finally, if any agent from \(\C\setminus {c_1}\) strongly envies \(c_1\), we replace \(c_1\)'s bundle with \(\mes_{c}(Y_{c_1},X_{c_2})\), which is still better than \(X_{c_1}\) for agent \(c_1\). Therefore, \(Y\) is an EFX allocation such that \(Y(L)\succ X(L)\) and \(Y(\N\setminus{L})=X(\N\setminus{L})\).\qedhere
\end{proof}
\section{Three Sources in \(E_X\)}\label{sec:three_sources}

As the leading agents hold the minimum valued bundle in their respective groups, they envy all the agents in their own group and are never envied by anyone in their own group. We consider the case when no leading agent is envied by anyone. That is, \(a_1,b_1\) and \(c_1\) are the sources in \(E_X\). 

% \textcolor{red}{Recall that $L = \{a_1, b_1, c_1\}$ denotes the set of the leading agents. For any set of agents $Q\subseteq \N$, and any allocation $X$, define $X(Q)=\{X_q\mid\ q\in Q\}$. Moreover, for two allocations $X,Y$, define $Y(Q)\succ X(Q)$ if $\forall q\in Q\ Y_q>_q X_q$. Also, we use $Y(Q)=X(Q)$ to denote $\forall q\in Q\ Y_q=_q X_q$.}

% Given an allocation \(X\), and a subset \(P\subseteq \N\) of agents, we define \(X(P)\) to be the allocation \(X\) with respect to the instance \(\langle P, \M, \rangle\)

\begin{restatable}{lemma}{threeSources} \label{lemma:three_sources}
    Let \(X\) be a partial EFX allocation with at least one unallocated good \(g\).  If the %set of 
    leading agents %in \(L = \{a_1,b_1,c_1\}\) 
    are sources in the envy graph \(E_X\),  then there exists an EFX allocation \(Y\) such that \(Y\succ X\). Furthermore, \(Y(L)\succ X(L)\) and \(Y(\N\setminus{L})=X(\N\setminus{L})\).
    %\textcolor{red}{PN: Furthermore, each agent in $L$ gets a more valuable bundle in $Y$ as compared to $X$ whereas the bundles of non-leading agents in $Y$ are same as those in $X$.}
\end{restatable}
% \begin{lemma}\label{lemma:three_sources}

\paragraph{Proof overview:} In this case, we only update the bundles of the leading agents, while the remaining agents retain their old bundles. After the update, each leading agent prefers their new bundle over their old ones. Additionally, the leading agents do not strongly envy the new bundles of other leading agents, even when compared to their own old bundles. Thus, the non-leading agents in a group (who keep their old bundles) do not strongly envy any agent in other groups. As a result, the only strong envies that remain are within the groups, which are resolved using Lemma \ref{lemma:internal_envies}. Therefore, the new allocation we obtain is an EFX allocation. Furthermore, the new bundles satisfy \(Y(L) \succ X(L)\) and \(Y(\mathcal{N} \setminus L) = X(\mathcal{N} \setminus L)\).

\begin{proof}[Proof of Lemma~\ref{lemma:three_sources}]
    From Proposition~\ref{prop:M_X_is_cyclic}, we know that the \(g\)-champion graph \(M_X\) must contain a cycle of length at most three, involving agents \(a_1, b_1\), and \(c_1\). If any of the leading agents self \(g\)-champions, using Corollary~\ref{corr:self_g_champion}, we get a Pareto-dominating EFX allocation \(Y\). If the \(g\)-champion cycle is of length \(2\), then by applying Lemma~\ref{lemma:mx_two_cycle}, we get our desired allocation. Therefore, assume that \(M_X\) has a cycle of length \(3\). Without loss of generality, assume that \((a_1,c_1)\),\((c_1,b_1)\) and \((b_1,a_1)\) are \(g\)-champion edges in \(M_X\). Since these agents do not envy each other, the \(g\)-champion edges break the bundle into top and bottom halves as shown below.
    \[ X(L) \ =\  \raisebox{-0.9em}{\hbox{\(\Allocation{T_{a_1}^{b_1}}{B_{a_1}^{b_1}}{T_{b_1}^{c_1}}{B_{b_1}^{c_1}}{T_{c_1}^{a_1}}{B_{c_1}^{a_1}}\)}} \]

    For ease of notation, we will call \(T_{a_1}^{b_1},T_{b_1}^{c_1}\), and \(T_{c_1}^{a_1}\) as \(T_{a},T_{b}\) and \(T_{c}\). Similarly, we call the bottom halves as \(B_{a},B_{b}\) and \(B_{c}\). We now make the following observations regarding the relationships among these parts of the bundles. 

    Firstly, agent \(a_1\) champions \(c_1\), but does not self \(g\)-champion. Therefore, \(T_c\cup g >_a T_a\cup g \implies T_c >_a T_a\). Similarly, as \(a_1\) does not \(g\)-champion \(b_1\), we have \(T_c\cup g >_a T_b\cup g \implies T_c >_a T_b\). 

    As \(T_c >_a T_a\), we have \(T_c \cup B_a >_a  T_a \cup B_a = X_{a_1}\). As \(a_1\) does not envy \(c_1\), we know that \(X_{a_1} = T_a\cup B_a >_a T_c\cup B_c\). By combining these two inequalities we get \(T_c \cup B_a >_a T_c\cup B_c \implies B_a>_a B_c\). 
    
    As \(a_1\) does not self \(g\)-champion, we have \(T_a\cup B_a >_a T_a\cup g \implies B_a >_a g\). As \(a_1\)  \(g\)-champions \(c_1\) but does not envy \(c_1\), we have \(T_c\cup g >_a T_c\cup B_c \implies g>_a B_c\). 

    Analogous observations can be made for agents \(b_1\) and \(c_1\). The following table summarizes the above observations for all three leading agents:

    \begin{align*}
        \text{Agent} \quad & \text{Observations} \\
        a_1: \quad & T_c >_a T_a; \quad T_c >_a T_b; \quad B_a >_a g >_a B_c  \tag{1}\label{condition:con1} \\
        b_1: \quad & T_a >_b T_b; \quad T_a >_b T_c; \quad B_b >_b g >_b B_a  \tag{2}\label{condition:con2} \\
        c_1: \quad & T_b >_c T_c; \quad T_b >_c T_a; \quad B_c >_c g >_c B_b  \tag{3}\label{condition:con3}
    \end{align*}
    
    Furthermore, as \(a_1\) does not envy \(b_1\), we have \(T_a\cup B_a >_a T_b\cup B_b\). Therefore, if \(B_a <_a B_b\), then it must be the case that \(T_a>_a T_b\). Combining with Condition~\ref{condition:con1}, if \(B_a <_a B_b\), then \(T_c>_a T_a>_a T_b\) and \(B_b>_a B_a>_a g >_a B_c\).  Analogous observations can be made for agents \(b_1\) and \(c_1\).
    
    We have three predicates: \(B_a <_a B_b, B_b <_b B_c\), and \(B_c <_c B_a\). We have the following four possibilities based of how many of these predicates hold true. 

    \begin{enumerate}[wide, labelindent=0pt,label=Case~\arabic*]

    \item\label{itm:all-hold} {\bf All the predicates hold:} That is, \(B_a <_a B_b, B_b <_b B_c\), and \(B_c <_c B_a\). We then have the following updated inequalities: 

    \begin{align*}
        \text{Agent} \quad & \text{Observations} \\
        a_1: \quad & T_c >_a T_a >_a T_b \quad\text{and}\quad  B_b >_a B_a >_a g >_a B_c  \tag{4}\label{condition:con4} \\
        b_1: \quad & T_a >_b T_b >_b T_c \quad\text{and}\quad  B_c >_b B_b >_b g >_b B_a  \tag{5}\label{condition:con5} \\
        c_1: \quad & T_b >_c T_c >_c T_a \quad\text{and}\quad  B_a >_c B_c >_c g >_c B_b  \tag{6}\label{condition:con6}
    \end{align*} 

    We allocate \(T_c \cup B_b\) to agent \(a_1\), \(T_a \cup B_c\) to agent \(b_1\), and \(T_b \cup B_a\) to agent \(c_1\). These bundles correspond to the most preferred top and bottom halves for each agent (refer to Conditions~\ref{condition:con4}, \ref{condition:con5}, and \ref{condition:con6}). Therefore, all three agents strictly improve. Furthermore, according to agent \(a_1\), we have \(T_c\cup B_b>_a T_a\cup B_a = X_{a_1} >_a T_a\cup B_c = Y_{b_1}\). That is, \(a_1\) does not envy \(b_1\) even with respect to her old bundle \(X_{a_1}\). Therefore, no agent in \(\A\) envy \(b_1\). Similarly, we have \(X_{a_1} >_a T_b\cup B_a = Y_{c_1}\). That is, no agent from \(\A\) envies \(c_1\). Similarly we can show that no agents from either \(\B\) or \(\C\) envy any of \(a_1,b_1\) or \(c_1\). If any strong envy remains, it must be confined to agents within their respective groups. Hence, we apply Lemma~\ref{lemma:internal_envies} to resolve these envies. \\

    \item\label{itm:none-hold} {\bf None of the three predicates holds:} Thus \(B_a >_a B_b\), \(B_b >_b B_c\), and \(B_c >_c B_a\). We then cyclically shift the top halves of these three bundles, allocating \(T_c \cup B_a\) to agent \(a_1\), \(T_a \cup B_b\) to agent \(b_1\), and \(T_b \cup B_c\) to agent \(c_1\).

    Now, since \(T_c >_a T_a\) we know that \(Y_{a_1} = T_c\cup B_a >_a T_a\cup B_a = X_{a_1}\). Therefore, agent \(a_1\) strictly improved in \(Y\). Similarly, we can show that agent \(b_1\) and agent \(c_1\) also strictly improves in \(Y\).

    If there is any strong envy in \(Y\), it must be directed towards or from the leading agents, as rest of the bundles remain unchanged. From case assumption, we have \(B_a >_a B_b\). Therefore \(X_{a_1} = T_a\cup B_a >_a T_a\cup B_b = Y_{b_1}\). As for each agent \(a_i\in \A\), \(Y_{a_i}\ge_a X_{a_1} >_a Y_{b_1}\), none of the agents in \(\A\) envy \(b_1\).

    Now we show that no agent in \(\A\) envies \(c_1\). Agent \(a_1\) did not envy \(c_1\) in \(X\). Therefore, \(X_{a_1}>_a T_{c}\cup B_{c}\). As \(T_c>_a T_b\) (Condition~\ref{condition:con1}), we have \(X_{a_1}>_a T_{c}\cup B_{c} >_a T_{b}\cup B_{c} = Y_{c_1}\). Therefore, \(a_1\) does not envy \(c_1\). Furthermore, as \(\forall a_i\in A,\ Y_{a_i}\ge_a X_{a_i}\), no agent in \(\A\) envies \(c_1\). 

    With a symmetric argument, we can show that no agents in \(\B\) or \(\C\) envies any of the leading agents. Therefore, \(Y\) is the required EFX allocation.

    % Since \(T_c >_a T_a\) and \(T_c >_a T_b\), agent \(a_1\) strictly improves and does not envy either \(b_1\) or \(c_1\) as they received less preferred top halves according to \(a_1\). Similarly, both \(b_1\) and \(c_1\) also experience improvements and do not envy any agents from \(\{a_1, b_1, c_1\}\). Furthermore, by applying the inequalities \ref{condition:con1},\ref{condition:con2}, and \ref{condition:con3} we have \(T_b>_a T_a\) which implies \(T_b\cup B_b>_a T_a\cup B_b = Y_{b_1}\). Along with the fact that \(a_1\) did not previously envy \(b_1\), we have  \(X_{a_1}>_a X_{b_1} >_a Y_{b_1}\). Therefore agent \(a_1\) does not envy \(b_1\) even with respect to her old bundle \(X_{a_1}\).  Therefore, no agent in \(\A\) envy \(b_1\). In an analogous way we can show that none of the agents envy any of \(a_1,b_1\) or \(c_1\). If any strong envy remains, it must be confined to agents within their respective groups. Hence, we apply Lemma~\ref{lemma:internal_envies} to resolve these envies.

    \item\label{itm:one-holds} {\bf One of the three predicates holds:} Without loss of generality, assume \(B_a <_a B_b\), \(B_b >_b B_c\), and \(B_c >_c B_a\).

    We then have:
    \begin{align*}
        \text{Agent} \quad & \text{Observations} \\
        a_1: \quad & T_c >_a T_a >_a T_b \quad\text{and}\quad  B_b >_a B_a >_a g >_a B_c  \tag{4}\label{condition:con7} \\
        b_1: \quad & T_a >_b T_b; \quad T_a >_b T_c; \quad B_b >_b g >_b B_a  \tag{2}\label{condition:con8} \\
        c_1: \quad & T_b >_c T_c; \quad T_b >_c T_a; \quad B_c >_c g >_c B_b  \tag{3}\label{condition:con9}
    \end{align*}

    We allocate the bundles as: \(T_c \cup B_a\) to agent \(a_1\), \(T_a \cup g\) to \(b_1\) and \(T_b\cup B_c\) to agent \(c_1\).

    Firstly, for agent \(a_1\), we have \(T_c \cup B_a >_a T_a \cup B_a\) and \(T_a \cup B_a >_a T_a \cup g\). Therefore, agent \(a_1\) strictly improves and does not envy \(b_1\) even with respect to her old bundle. Thus no agent in \(\A\) envy \(b_1\). Furthermore, since \(T_c \cup B_a >_a T_a \cup B_a >_a T_b \cup B_c\), neither agent  \(a_1\) nor any other agent from \(\A\) envies \(c_1\).

    Secondly, as \(T_a \cup g\) is the \(\mes\) bundle, agent \(b_1\) strictly improves, i.e., \(T_a \cup g >_b T_b \cup B_b\). Given the case assumption \(B_b >_b B_c\), we have \(T_b \cup B_b >_b T_b \cup B_c\). Thus, none of the agents from \(\B\) envy \(c_1\).   As \(b_1\) did not \(g\)-champion \(c_1\), we have \(T_b\cup B_b >_b T_c \cup g\). From Condition~\ref{condition:con8}, we know that \(T_c \cup g >_b T_c \cup B_a\). Therefore, as \(T_b\cup B_b >_b T_c \cup B_a\), no agent in \(\B\) envies \(a_1\). 

    Finally, As \(T_b >_c T_c\), agent \(c_1\) also has strict improvement. That is, \(T_b \cup B_c >_c T_c \cup B_c\). Because of the case assumption \(B_c >_c B_a\), we have  \(T_c\cup B_c >_c T_c\cup B_a\). As no agent in \(\C\) previously envied \(X_{c_1}=T_c\cup B_c\), no agent in \(\C\) envies \(a_1\) in \(Y\). As \(c_1\) or any other agent in \(\C\) did not previously \(g\)-champion \(a_1\), they do not envy \(b_1\) who has the bundle \(T_a \cup g\). 

    Therefore, if any strong envy remains, it must be confined to agents within their respective groups. Hence, we apply Lemma~\ref{lemma:internal_envies} to resolve these envies.

    \item\label{itm:two-hold} {\bf Two out of the three predicates hold true:} Without loss of generality, assume \(B_a <_a B_b\), \(B_b <_b B_c\), and \(B_c >_c B_a\). We have the following observations: 

    \begin{align*}
        \text{Agent} \quad & \text{Observations} \\
        a_1: \quad & T_c >_a T_a >_a T_b \quad\text{and}\quad  B_b >_a B_a >_a g >_a B_c  \tag{4}\label{condition:con4.1} \\
        b_1: \quad & T_a >_b T_b >_b T_c \quad\text{and}\quad  B_c >_b B_b >_b g >_b B_a  \tag{5}\label{condition:con4.2}
    \end{align*} 

    \begin{enumerate}[wide,  labelindent=0pt,label*=.\arabic*]
    \item\label{itm:c>a} {\(\mathbf{T_c >_c T_a}\):} Now, along with the above two observations, we have 
    \begin{align*}
        c_1: \quad & T_b >_c T_c >_c T_a \quad\text{and}\quad  B_c >_c g >_c B_b \quad\text{and}\quad B_c>_c B_a \tag{6}\label{condition:con4.3}
    \end{align*}

    We allocate \(T_c \cup B_b\) to \(a_1\), \(T_a \cup B_c\) to \(b_1\) and \(T_b \cup g\) to agent \(c_1\).

    For agents \(a_1\) and \(b_1\), it is clear from Conditions~\ref{condition:con4.1} and \ref{condition:con4.2} that they have received strict improvements with respect to their previous bundles and that they do not envy anyone in \(\{a_1, b_1, c_1\}\) in the new allocation. Furthermore, from \ref{condition:con4.1} we see that  \(T_a\cup B_a >_a T_a\cup B_c = Y_{b_1}\) and \(T_a\cup B_a >_a T_b\cup g = Y_{c_1}\). Therefore, no agent from \(\A\) envies either \(b_1\) or \(c_1\). Similarly, using \ref{condition:con4.2} we can show that agents in \(\B\) envy neither \(a_1\) nor \(c_1\). 
    
    For agent \(c_1\), we know that \(T_b\cup g >_c T_c \cup B_c\) as \(c_1\) \(g\)-championed \(b_1\). By applying Condition~\ref{condition:con4.3}, we get \(T_c \cup B_c >_c T_c \cup B_b\) and  \(T_c \cup B_c >_c T_a \cup B_c\). Therefore, agent \(c_1\) (or any other agent in \(\C\)) envies neither \(b_1\) nor \(a_1\). Therefore, any remaining strong envy must be restricted to agents within their respective groups. Consequently, we apply Lemma~\ref{lemma:internal_envies} to resolve these envies.

    \item\label{itm:c<a} {\(\mathbf{T_c <_c T_a}\):} Therefore, instead of Condition~\ref{condition:con4.3}, we have the following observation: 

    \begin{align*}
        c_1: \quad & T_b >_c T_a >_c T_c \text{ and } B_c >_c g >_c B_b \text{ and } B_c>_c B_a \tag{7}\label{condition:con4.4}
    \end{align*}

    In this case, we would like to allocate \(T_c\cup B_a\) to \(a_1\), \(T_b\cup B_b\) to \(b_1\) and \(T_a\cup B_c\) to agent \(c_1\). %\nv{I am confused by the above potential allocation -- in the line preceding this comment. I can't see why that allocation would have been ideal.}\vp{(vp: removed all that confusing lines.)} H
    However, in this allocation agent \(b_1\) envies agent \(c_1\), and this envy could potentially be a strong envy. Therefore, we consider the \(\mes\) bundle \(S =\mes_b(T_a\cup B_c, X_{b_1})\). Based on whether or not agent \(c_1\) prefers \(S\) over \(X_{c_1}\), we have the following two sub-cases:\\
    
    \begin{enumerate}[wide, labelindent=0pt,label*=(\alph*)]
        
        \item\label{itm:S>Xc1} { \(\mathbf{S >_c X_{c_1}}\):} Then, we allocate the bundles as follows. Agent \(a_1\) gets \(T_c \cup B_a\), \(b_1\) retains \(T_b\cup B_b\) and \(c_1\) gets \(S\). From Condition~\ref{condition:con4.1} we have \(T_c\cup B_a >_a T_a\cup B_a\) and from this case assumption we know that \(S>_c X_{c_1}\). Therefore agents \(a_1\) and \(c_1\) receive strict improvements while agent \(b_1\) retains her old bundle.

    We know that \(X_{a_1} >_a X_{b_1} = Y_{b_1}\) Therefore, not agent from \(\A\) envies \(b_1\). Furthermore, as \(B_a >_a B_c\) (Condition~\ref{condition:con4.1}), we get \(T_a \cup B_a >_a T_a\cup B_c >_a S\). Thus, no agent from \(\A\) envies \(c_1\). 

    The bundle \(S\) is an \(\mes\) bundle of agent \(b_1\). Therefore, agent \(b_1\) (or any other agent in \(\B\)) does not strongly envy \(c_1\). From Condition~\ref{condition:con4.2}, we have \(T_a >_b T_c\). Therefore, \(T_a \cup B_a >_b T_c \cup B_a\). Since \(b_1\) did not previously envy \(a_1\), we also have \(T_b\cup B_b >_b T_a \cup B_a\). By combining these two inequalities, we get \(T_b\cup B_b >_b T_c \cup B_a \). Hence, \(b_1\) (and therefore no one else from \(\B\)) does not envy \(a_1\).

    We know that \(S>_c X_{c_1} = T_c\cup B_c\). From Case~\(4\) assumption we have \(B_c >_c B_a\). Therefore, \(S>_c T_c\cup B_c >_a T_c\cup B_a\). Therefore, none of the agents in \(\C\) envy \(a_1\). As \(c_1\) or any other agent in \(\C\) did not previously envy \(b_1\), they do not envy \(b_1\) now. 

    Therefore, any remaining strong envy must be restricted to agents within their respective groups. Consequently, we apply Lemma~\ref{lemma:internal_envies} to resolve these envies.\\
    
    \item\label{itm:S<Xc1}{\(\mathbf{S <_c X_{c_1}}\):} Then, we allocate the bundles as follows:  Agent \(a_1\) gets \(T_c \cup B_b\), \(b_1\) gets \(S\) and \(c_1\) gets her \(\mes\) bundle \(T_b\cup g\). 

    According to agents in  \(\A\): Applying Condition~\ref{condition:con4.1}, \(T_c\cup B_b >_a T_a\cup B_a >_a T_a\cup B_c >_a S\). Therefore, \(a_1\)'s bundle strictly improved and \(a_1\) does not envy \(b_1\) even with respect to her old bundle. Thus no agent in \(\A\) envy \(b_1\). Since \(a_1\) or anyone else from \(\A\) did not previously \(g\)-champion \(b_1\), They do not envy \(c_1\) who has the bundle \(T_b\cup g\). 

    According to agents in \(\B\): As \(S\) is \(b_1\)'s \(\mes\) bundle, \(b_1\) received a strictly better bundle. From Condition~\ref{condition:con4.2}, we know that \(T_b >_b T_c\), therefore, \(S >_b T_b\cup B_b >_b T_c \cup B_b\). That is, \(b_1\) does not envy \(a_1\), even with respect to her old bundle. Thus, no agent in \(\B\) envy \(a_1\). Since \(b_1\) did not previously self \(g\)-champion, and none of the agents in \(\B\) \(g\)-championed \(b_1\), they do not envy \(T_b\cup g\) of agent \(c_1\). 

    According to agents in  \(\C\): \(T_b\cup g\) is an \(\mes\) bundle of agent \(c_1\). Therefore, \(c_1\) received strict improvement. Using Condition~\ref{condition:con4.4}, we know \(B_c >_c B_b\). Therefore, \(T_c \cup B_c >_c T_c\cup B_b\). That is, \(c_1\) does not envy \(a_1\), even with respect to her old bundle. Thus no agent in \(\C\) envies \(a_1\) . From the current case assumption, that is \(S<_c X_{c_1}\), we know that no agent in \(\C\) envies \(b_1\). 

    Therefore, the current allocation Pareto dominates \(X\)  and any strong envy that remains must be restricted to agents within their respective groups. Thus, we apply Lemma~\ref{lemma:internal_envies} to resolve these envies. \qedhere
    \end{enumerate}
    \end{enumerate}
    \end{enumerate}
\end{proof}

\begin{corollary}\label{cor:three_source}
    Let \(X\) be a partial EFX allocation with at least one unallocated good \(g\). If the leading agents in \(L = \{a_1,b_1,c_1\}\) are all the sources in the envy graph \(E_X\), then there exists an EFX allocation \(Y\), possibly partial, such that at least one of the following holds:
    \begin{enumerate}
        \item \(\phi(Y)>\phi(X)\)
        \item \(Y\succ X\) and \(a_1\) is a source in \(E_Y\) and \(\forall \alpha\in (\B\cup\C) \setminus L\) if \((a_1,\alpha)\) is an edge in  \(E_X\) then \((a_1,\alpha)\) is an edge in \(E_Y\).
    \end{enumerate}
\end{corollary}
\begin{proof}
    Given a partial EFX allocation \(X\) with three sources, we apply Lemma~\ref{lemma:three_sources} to get an allocation \(Y\) such that \(Y(L)\succ X(L)\) and \(Y(\N\setminus{L})=X(\N\setminus{L})\). If agent \(a_1\) has strictly improved in \(Y\), then condition \(1\) is satisfied. 

    Otherwise, agent \(a_1\) must have retained the same bundle in \(Y\) as in \(X\). As no agent envied \(a_1\) in \(X\) who had the same bundle as in \(Y\), they continue to not envy \(a_1\) in \(Y\) as  \(Y\setminus a_1\succ X\setminus a_1\). Therefore \(a_1\) is a source in \(Y\). Additionally, if agent \(a_1\) envied some agent in \((\B\cup\C)\setminus L\) in \(X\), then \(a_1\) continues to envy the same agent in \(Y\) as \(Y(\N\setminus L) = X(\N \setminus L)\). \qedhere
\end{proof}

\section{Agent \(a_1\) Envies someone in \(\B\cup \C\)}\label{sec:a_1_envies_someone}

In this section we consider the case when agent \(a_1\) is a source and envies some agent from \(\B\cup \C\). 

If \(a_1\) envies either \(b_1\) or \(c_1\), we have the following lemma.

\begin{restatable}{lemma}{aEnviesborc}
\label{lemma:a_1envies_b_1_or_c_1}
    Let \(X\) be a partial EFX allocation with at least one unallocated good \(g\). If \(a_1\) is a source in \(E_X\) and \(a_1\) envies either \(b_1\) or \(c_1\), then there exists an EFX allocation \(Y\) such that \(\phi(Y)>\phi(X)\).
\end{restatable}
\begin{proof}
    Without loss of generality, assume that \(a_1\) envies \(b_1\). If \(a_1\) is the only source in \(E_X\), we immediately get our desired allocation \(Y\) by applying Corollary~\ref{corr:single_source_phi_improvement}. The remaining case is when there are two sources in \(E_X\). Thus, \(c_1\) is also a source in \(E_X\). Given that some agent must \(g\)-champion \(a_1\), we first consider the case where \(a_1\) self \(g\)-champions. In this case, we can improve \(a_1\)'s bundle by replacing \(X_{a_1}\) with the \(\mes\) bundle \(T_{a_1}^{a_1} \cup g\). We get \(Y\), \(\phi(Y)>\phi(X)\).

    If, however, \(b_1\) \(g\)-champions \(a_1\), then \(\langle (a_1,b_1),(b_1,a_1)\rangle\) is a pseudo-cycle, thus we are done. The only remaining possibility is that \(c_1\) \(g\)-champions \(a_1\). We now consider cases based on who \(g\)-champions \(c_1\).

    \begin{enumerate}[wide,  labelindent=0pt,label=Case~\arabic*]
        \item\label{itm:a1-champion-c1} {\bf Agent \(a_1\) \(g\)-champions \(c_1\):} Then \(a_1\) and \(c_1\) form a \(g\)-champion cycle. Note that none of the leading agents envy $a_1$ or $c_1$ as they are both sources. In this case, we can apply Lemma~\ref{lemma:mx_two_cycle} to obtain an allocation \(Y\) such that \(\phi(Y) > \phi(X)\).

        \item\label{itm:c1-self-champion} {\bf Agent \(c_1\) self \(g\)-champions: } We compute a new Pareto-dominating EFX allocation by replacing \(X_{c_1}\) with the \(\mes\) bundle \(T_{c_1}^{c_1} \cup g\). In this new allocation, \(a_1\) continues to be a source and still envies \(b_1\). We start again by applying Lemma~\ref{lemma:a_1envies_b_1_or_c_1} to this new allocation \(Y\). Since Pareto dominance is a partial order and there are finitely many partial EFX allocations, this procedure must terminate with an allocation \(Z\) such that either \(Z\) is a complete EFX allocation, or \(\phi(Z) > \phi(X)\).

        \item\label{itm:b1-champion-c1} {\bf Agent \(b_1\) \(g\)-champions \(c_1\):} See figure below for an illustration of this case.
        
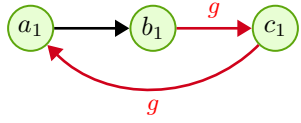
\begin{center}
        %\begin{figure}
            \begin{tikzpicture}
                % Nodes
                \node[leading_agent] (a1) {\small \(a_1\) };
                \node[leading_agent, right=1cm of a1] (b1) {\small \(b_1\) };
                \node[leading_agent, right=1cm of b1] (c1) {\small \(c_1\) };
            
                % Edges
                \draw[envy] (a1) -- (b1);
                \draw[champ] (b1) to[out=0, in=180] node[above, red] {\footnotesize \(g\)} (c1);
                \draw[champ] (c1) to[out=45, in=135] node[below, red] {\footnotesize \(g\)} (a1);

                % \draw[champ, loop below, min distance=10mm] (b1.north east) to[in=-55, out=235] node[above, red] {\footnotesize \(g\)} (b1.north west);
            \end{tikzpicture}
            %\label{fig:b1-champion-c1}
            %\end{figure}
        \end{center}

        We consider a new allocation by exchanging the top halves as follows:
        \[ Y\ =\  \raisebox{-0.9em}{\hbox{\(\Allocationfiveone{X_{b_1}}{T_{c_1}^{b_1}}{g}{T_{a_1}^{c_1}}{B_{c_1}^{b_1}}\)}} \]
        Observe that the bundles of all the three agents \(a_1,b_1\), and \(c_1\) have strictly improved and therefore \(\phi(Y)>\phi(X)\). We now check if the new allocation is indeed EFX. If there are no strong envies in \(Y\), we are done. So, consider the case when there is at least one strong envy in \(Y\). Observer that if there are any strong envies in \(Y\), they must be towards either \(a_1\) or \(b_1\) or \(c_1\). We know that no agent strongly envies \(a_1\) or \(b_1\) as \(a_1\) received an old bundle from \(X\) and \(Y_{b_1}\) was an \(\mes\) bundle in \(X\). Furthermore, agent \(b_1\) does not envy \(c_1\). This is because, from Proposition~\ref{prop:i_like_the_top_that_i_cut} we have \(T_{c_1}^{b_1} >_b T_{a_1}^{c_1}\). Therefore, \(X_{c_1} = T_{c_1}^{b_1} \cup B_{c_1}^{b_1} >_b T_{a_1}^{c_1} \cup B_{c_1}^{b_1}\). Since \(b_1\) did not envy \(X_{c_1}\) in \(X\), she does not envy \(T_{a_1}^{c_1} \cup B_{c_1}^{b_1}\) in \(Y\). Thus, the only possible strong envy is from agent \(a_1\) or other agents in \(\A\), towards \(c_1\). We know that \(c_1\) does not envy either \(a_1\) or \(b_1\). However, \(c_1\) might envy one of the other agents. We consider the following cases based on possible envies from agent \(c_1\) in the allocation \(Y\).

            \begin{enumerate}[wide,  labelindent=0pt,label*=.\arabic*]
                \item\label{itm:c1-envies-ai} {\bf Agent \(c_1\) envies some \(a_i\in \A\setminus\{a_1\}\):} Then, we exchange the bundles of agents \(a_i\) and \(c_1\). Note that there may not be envy from \(a_i\) to \(c_1\), But since $a_1$ envies $c_1$, the bundle of $c_1$ is better than bundle of $a_1$ according to $v_a$. Now, all the strong envies are within the groups. Therefore, we apply Corollary~\ref{corr:intra-group} to resolve such envies and obtain a new EFX allocation \(Z\) such that \(\phi(Z)=\phi(Y)>\phi(X)\). 

                \item\label{itm:c1-envies-bj} {\bf Agent \(c_1\) envies some \(b_j \in \B \setminus \{b_1\}\):} In this case, we reallocate as follows: assign \(Y_{b_j}\) to \(c_1\), \(Y_{c_1}\) to \(a_1\), and \(Y_{a_1}\) (which is same as \(X_{b_1}\)) back to \(b_1\). As \(a_1\) received a bundle which was strongly envied by some agent in \(\A\), \(a_1\)'s new bundle is strictly better than \(X_{a_1}\). Since \(c_1\) received the bundle that she envied, she also sees a strict improvement. 
         
                However, the value of \(b_1\)'s bundle decreases. Recall that in the previous allocation \(X\), we established that \(X_{c_1} >_b T_{a_1}^{c_1} \cup B_{c_1}^{b_1}\), therefore \(b_1\) does not envy \(a_1\). Furthermore, since \(Y_{b_1}\) was an \(\mes\) bundle relative to \(X_{b_1}\), agent \(b_1\) does not \emph{strongly} envy \(c_1\). At this point, all remaining strong envies are confined within the groups. Therefore, we apply Lemma~\ref{lemma:internal_envies} to resolve these envies, yielding a new EFX allocation \(Z\) such that \(\phi(Z) = \phi(Y) > \phi(X)\).

                \item\label{itm:c1-envies-none} {\bf Agent \(c_1\) does not envy anyone from \(\A \cup \B\):} Let \(S \subseteq Y_{c_1}\) denote the smallest subset of \(Y_{c_1}\) such that, according to \(c_1\)'s valuation, \(S\) is preferred over the bundles of all agents in \(\A \cup \B\). If \(a_1\) does not strongly envy \(S\), then we allocate \(S\) to \(c_1\) and resolve any internal strong envies. This gives us the desired EFX allocation which \(\phi\)-dominates the previous one. 
        
                Suppose \(a_1\) strongly envies \(c_1\) even with the bundle \(S\). Then, let \(S'\subset S\) be a smallest cardinality subset of \(S\) that \(a_1\) \emph{weakly} envies. Let \(T=\max_{c}\{X_{p}\mid p\in \A \cup \B\}\) be the best bundle among the agents in \(\A\cup \B\) according to agent \(c_1\). \(T\) can either be a bundle from one of the agents in \(\A\) or from \(\B\). 
                    \begin{enumerate}[wide,  labelindent=0pt,label*=.\arabic*]
                        \item\label{itm:T=Yai} {\bf \(\mathbf{T=Y_{a_i}}\) for some \(\mathbf{a_i\in \A}\):} Then allocate \(T\) to \(c_1\) and \(S'\) to \(a_i\).  We then resolve internal strong envies within the groups using Corollary~\ref{corr:intra-group}, yielding a new EFX allocation \(Z\) such that \(\phi(Z) = \phi(Y) > \phi(X)\).

                        \item\label{itm:T=Ybj} {\bf \(\mathbf{T = Y_{b_j}}\) for some \(\mathbf{b_j \in \B}\):} In this case, we ask agent \(b_j\) to select her most preferred bundle from the agents in \(\A\), say from agent \(a_i\). We then allocate \(Y_{a_i}\) to \(b_j\), \(Y_{b_j}\) to \(c_1\), and \(S'\) to agent \(a_i\). We know that agent \(b_1\) does not strongly envy any agent from \(\A \cup \C\). Thus, we swap the bundles of \(b_j\) and \(b_1\), so that now agent \(b_j\) does not strongly envy anyone from \(\A \cup \C\). Consequently, any strong envy towards agents in \(\A \cup \C\), if it exists, must now come from agent \(b_1\).

                        In this allocation, agent \(c_1\) does not envy any agent from \(\A \cup \B\) because she chose her most preferred bundle. Agent \(a_1\)'s prior strong envy towards \(c_1\) is now resolved. Additionally, agent \(a_1\) does not strongly envy \(b_1\), as she did not strongly envy \(a_i\) earlier. Therefore, \(a_1\) does not strongly envy any agent from \(\B \cup \C\).

                        Finally, agent \(b_1\) does not envy any agent in \(\A \setminus \{a_i\}\), as \(b_j\) (who has the same valuation as \(b_1\)) selected the best bundle from \(\A\). The bundle \(Y_{a_i}\), which \(b_j\) picked, is at least as valuable as \(Y_{a_1} = X_{b_1}\). Moreover, recall that with bundle \(X_{b_1}\), agent \(b_1\) did not strongly envy \(b_j\) or any agent in \(\C\) in allocation \(X\). Therefore, in the new allocation, with a bundle at least as valuable as \(X_{b_1}\), agent \(b_1\) does not strongly envy any agent in \(\C\). Additionally, we know that \(X_{c_1} >_b T_{a_1}^{c_1} \cup B_{c_1}^{b_1} >_b S'\), so agent \(b_1\) does not strongly envy \(a_i\).

                        Based on the above observations, agent \(b_1\) does not strongly envy any agent from \(\B \cup \A\). Therefore, we apply Corollary~\ref{corr:intra-group} to resolve any internal strong envies, resulting in a new EFX allocation \(Z\) such that \(\phi(Z) = \phi(Y) > \phi(X)\).\qedhere
                    \end{enumerate}
            \end{enumerate}
    \end{enumerate} 
\end{proof}

\begin{lemma}\label{lemma:a_1_envies_somebody}
     Let \(X\) be a partial EFX allocation with at least one unallocated good \(g\). If agent \(a_1\) is a source in \(E_X\) and if \(a_1\) envies some agent from \(\B\cup\C\), then there exists an EFX allocation \(Y\) such \(\phi(Y)>\phi(X)\).
\end{lemma}
\begin{proof}
    Agent \(a_1\) envies some agent in \(\B\cup\C\). If \(a_1\) envies either \(b_1\) or \(c_1\), we can apply Lemma~\ref{lemma:a_1envies_b_1_or_c_1} to get an allocation \(Y\) such that \(\phi(Y)>\phi(X)\). Thus, for the remainder of this proof, we consider the case where agent \(a_1\) envies some agent in \((\B\cup\C)\setminus\{b_1,c_1\}\).
    
    We now consider various cases based on the number of sources in the envy graph \(E_X\).
\begin{enumerate}[wide,  labelindent=0pt,label=Case~\arabic*]
    \item\label{itm:3-sources} {\bf \(E_X\) has three sources: }
    From Proposition~\ref{prop:M_X_is_cyclic}, we know that none other than leading agents \(L=\{a_1,b_1,c_1\}\) are the sources in \(E_X\). We apply Corollary~\ref{cor:three_source} to get an allocation \(Y\). If \(\phi(Y)>\phi(X)\), we are done. Otherwise, because of Condition~\(2\) of Corollary~\ref{cor:three_source}, we know that \(Y\succ X\) and \(Y\) satisfies the conditions required to apply Lemma~\ref{lemma:a_1_envies_somebody}. So, we apply Lemma~\ref{lemma:a_1_envies_somebody} on \(Y\). Since Pareto dominance is a partial order and there are finitely many partial EFX allocations, this procedure must terminate with an allocation \(Z\) such that either \(Z\) is a complete EFX allocation or \(\phi(Z)>\phi(X)\).

    \item\label{itm:a1-single-source} {\bf Agent \(a_1\) is a single source in \(E_X\):} Then by applying Lemma~\ref{lemma:kavita}, we get an allocation \(Y\) such that \(\phi(Y)>\phi(X)\).

    \item\label{itm:2-sources} {\bf \(E_X\) has two sources:} Without loss of generality, assume that \(a_1\) envies some agent \(b_i\in \B\setminus\{b_1\}\). Furthermore, w.l.o.g assume that there are no self \(g\)-champions. We know that some agent must \(g\)-champion \(a_1\).

\begin{enumerate}[wide,  labelindent=0pt, label*=.\arabic*]
    \item\label{itm:b1-champion-a1} {\bf Agent \(b_1\) \(g\)-champions \(a_1\):} Then, \(\langle(a_1,b_i),(b_1,a_1)\rangle\) forms a pseudo-cycle. Therefore, using Lemma~\ref{lemma:pseudo-cycle}, we get a new EFX allocation \(Y\) such that \(\phi(Y)>\phi(X)\).

    \item\label{itm:c1-champion-a1} {\bf Agent \(c_1\) \(g\)-champions \(a_1\):} Recall that both \(b_1\) and \(c_1\) cannot be sources in \(E_x\) because of Case~\(3\) assumption. Furthermore, we have assumed that \(a_1\) envies neither \(b_1\) nor \(c_1\). Therefore, either \(b_1\) envies \(c_1\) or \(c_1\) envies \(b_1\).

\begin{enumerate}[wide,  labelindent=0pt, label*=.\arabic*]
    \item\label{itm:b1-envies-c1} {\bf Agent \(b_1\) envies \(c_1\):} Here \(\langle(a_1,b_i),(b_1,c_1),(c_1,a_1)\rangle\) is a pseudo-cycle. Therefore, using Lemma~\ref{lemma:pseudo-cycle}, we get a new EFX allocation \(Y\) such that \(\phi(Y)>\phi(X)\).

     \item\label{itm:c1-envies-b1} {\bf Agent \(c_1\) envies \(b_1\):} In this case, we know that someone must \(g\)-champion agent \(c_1\).
\begin{enumerate}[wide,  labelindent=0pt, label*=.\arabic*]     
     \item\label{itm:b1-champion-c1_in_case_3} {\bf \(b_1\) champions \(c_1\):} we construct a new EFX allocation \(Y\) by giving the bundle \(X_{b_1}\) to \(c_1\) and \(T_{c_1}^{b_1}\cup g\) to agent \(b_1\). Note that both agent \(b_1\) and \(c_1\) have strictly improved. Furthermore, as \(T_{c_1}^{b_1}\cup g\) is a minimally envied subset, no agent strongly envies \(b_1\). As \(X_{b_1}\) was not strongly envied in \(X\), it cannot be strongly envied in \(Y\). Therefore, \(Y\) is an EFX allocation such that \(Y\succ X\). Also note that agent \(a_1\) continues to envy agent \(b_i\). Thus, allocation \(Y\) satisfies the conditions required to apply Lemma~\ref{lemma:a_1_envies_somebody} and \(Y\succ X\). So, we apply Lemma~\ref{lemma:a_1_envies_somebody} on \(Y\). Since Pareto dominance is a partial order and there are finitely many partial EFX allocations, this iterative procedure must terminate with an allocation \(Z\) such that either \(Z\) is a complete EFX allocation or \(\phi(Z)>\phi(X)\).

      \item\label{itm:a1-champion-c1_in_case_3} {\bf Agent \(a_1\) \(g\)-champions \(c_1\):} Then, we exchange the top halves of \(X_{a_1}\) and \(X_{c_1}\) to get a new allocation \(Y\), where the bundles of \(a_1,b_1\) and \(c_1\) are as follows:

      \[ Y(L) \ =\  \raisebox{-0.9em}{\hbox{\(\Allocationfivetwo{T_{c_1}^{a_1}}{B_{a_1}^{c_1}}{X_{b_1}}{T_{a_1}^{c_1}}{B_{c_1}^{a_1}}\)}} \]

      Applying Proposition~\ref{prop:i_like_the_top_that_i_cut}, we know that \(T_{c_1}^{a_1}>_{a} T_{a_1}^{c_1}\). Therefore, \(Y_{a_1}>_{a}X_{a_1}>_{a}Y_{c_1}\). That is, agent \(a_1\) has strictly improved and no agent from \(\A\) strongly envies \(c_1\) in \(Y\). Analogously, \(Y_{c_1}>_{c}X_{c_1}>_{c}Y_{a_1}\). That is, agent \(c_1\) has strictly improved and no agent from \(\C\) strongly envies \(a_1\) in \(Y\). Since agent \(b_1\) was envying neither \(a_1\) nor \(c_1\) in \(X\), \(b_1\) can envy at most one of \(a_1,c_1\) in \(Y\). Without loss of generality, assume that \(b_1\) strongly envies \(a_1\). We then replace the bottom half \(B_{a_1}^{c_1}\) of \(a_1\) with \(g\) to get \(Z\). The bundles of \(a_1,b_1,\) and \(c_1\) in \(Z\) are as follows:

      \[ Z(L) \ =\  \raisebox{-0.9em}{\hbox{\(\Allocationfivetwo{T_{c_1}^{a_1}}{g}{X_{b_1}}{T_{a_1}^{c_1}}{B_{c_1}^{a_1}}\)}} \]

      As \(T_{c_1}^{a_1}\cup g\) is a minimally envied subset, no one strongly envies agent \(a_1\). Furthermore, \(Z_{a_1}>_a X_{a_1}>_a Z_{c_1}\), therefore agent \(a_1\) is strictly better off compared to \(X\) and \(a_1\) does not strongly envy \(c_1\). Therefore, we have an EFX allocation \(Z\) such that \(\phi(Z)>\phi(X)\). \qedhere
      \end{enumerate}
      \end{enumerate}
      \end{enumerate}
      \end{enumerate}
\end{proof}

\section{Agent \(a_1\) Envies no one in \(\B\cup \C\)}\label{sec:a_1_envies_none}
Here we consider a partial EFX allocation $X$, where agent $a_1$ is a source and $a_1$ does not envy anyone in $\B\cup \C$.

\begin{restatable}{lemma}{aenviesnone}\label{lemma:a_1_envies_none}
    Let \(X\) be a partial EFX allocation with at least one unallocated good \(g\). If \(a_1\) is a source in \(E_X\) and \(a_1\) envies no agent from \(\B\cup\C\), then there exists an EFX allocation \(Y\) such \(\phi(Y)>\phi(X)\).
\end{restatable}
\begin{proof}
    We consider all possible cases based on the number of sources in \(E_X\).
    
    \begin{enumerate}[wide,  labelindent=0pt, label= \textbf{Case~\arabic*.}]
        \item If \(a_1\) is the only source in \(E_X\), we apply Corollary~\ref{corr:single_source_phi_improvement} to get an EFX allocation \(Y\) such that \(\phi(Y)>\phi(X)\).

        \item Suppose there are three sources in \(E_X\). That is, \(L=\{a_1,b_1,c_1\}\) is the set of sources in \(E_X\). We apply Corollary~\ref{cor:three_source} on \(X\) to get an allocation \(Y\), \(Y\succ X\). If \(\phi(Y)>\phi(X)\), we are done. Otherwise, we know that \(a_1\) is a source in \(Y\). If \(a_1\) envies some agent in \(\B\cup\C\), we apply Lemma~\ref{lemma:a_1_envies_somebody} on \(Y\) to get an allocation \(Z\), such that \(\phi(Z)>\phi(Y)=\phi(X)\). If \(a_1\) envies no agent in \(\B\cup \C\), we apply Lemma~\ref{lemma:a_1_envies_none} on \(Y\), as \(Y\) satisfies all the conditions to apply Lemma~\ref{lemma:a_1_envies_none} and \(Y\succ X\). Since Pareto dominance is a partial order and there are finitely many partial EFX allocations, this iterative procedure must terminate with an EFX allocation \(Z\) such that either \(Z\) is a complete allocation or \(\phi(Z)>\phi(X)\).

        \item There are two sources in \(E_X\).
        We know that \(a_1\) is a source. Without loss of generality, let \(b_1\) be the other source. Therefore \(c_1\) is not a source and hence it must be envied by some leading agent. Since \(a_1\) does not envy anyone, \(b_1\) envies \(c_1\). We now consider the cases depending on which agent \(g\)-champions \(b_1\). 

            \begin{enumerate}[wide,  labelindent=0pt, label*=\textbf{\arabic*.}]
                \item Suppose agent \(b_1\) self \(g\)-champions (see the figure below). We then construct a new Pareto dominating EFX allocation \(Y\) by replacing \(X_{b_1}\) with \(\mes_b(X_{b_1}\cup g)\). Since this is a strict improvement for agent \(b_1\), she continues to not envy \(a_1\) even in \(Y\). No one else envies \(a_1\) as there was no change in their bundles. Thus \(a_1\) is a source in \(Y\). If \(a_1\) envies \(b_1\), we apply Lemma~\ref{lemma:a_1_envies_somebody} to get a \(\phi\) improvement. Otherwise, we apply Lemma~\ref{lemma:a_1_envies_none} again on \(Y\). As there are finitely many EFX allocations, this Pareto improvement cannot go on for ever.
    
                
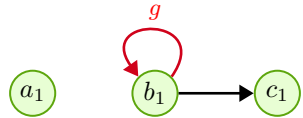
\begin{center}
                    \begin{tikzpicture}
                        % Nodes
                        \node[leading_agent] (a1) {\small \(a_1\) };
                        \node[leading_agent, right=1cm of a1] (b1) {\small \(b_1\) };
                        \node[leading_agent, right=1cm of b1] (c1) {\small \(c_1\) };
                    
                        % Edges
                        \draw[envy] (b1) -- (c1); 
                        \draw[champ, loop below, min distance=10mm] (b1.north east) to[in=-55, out=235] node[above, red] {\footnotesize \(g\)} (b1.north west);
                    \end{tikzpicture}
                \end{center}
    
                \item If \(c_1\) \(g\)-champions \(b_1\) (See the figure below), then \((b_1,c_1,b_1)\) is a Pareto improving cycle. Eliminating this cycle gives an allocation \(Y\) such that \(Y\succ X\) and \(a_1\) is a source in \(E_Y\) and \(a_1\) does not envy any agent in \(\B\cup\C\). We apply Lemma~\ref{lemma:a_1_envies_none} again on \(Y\). As there are finitely many EFX allocations, this Pareto improvement cannot go on for ever. 

                
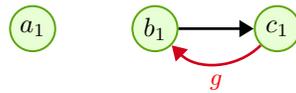
\begin{center}
                    \begin{tikzpicture}
                        % Nodes
                        \node[leading_agent] (a1) {\small \(a_1\) };
                        \node[leading_agent, right=1cm of a1] (b1) {\small \(b_1\) };
                        \node[leading_agent, right=1cm of b1] (c1) {\small \(c_1\) };
                    
                        % Edges
                        \draw[envy] (b1) -- (c1); 
                        \draw[champ] (c1) to[out=45, in=135] node[below, red] {\footnotesize \(g\)} (b1);
                    \end{tikzpicture}
                \end{center}

                \item If \(a_1\) \(g\)-champions \(b_1\), then we give an algorithm (Algorithm~\ref{alg:multi-competition}) to get an allocation \(Y\) such that \(\phi(Y)>\phi(X)\). Lemma~\ref{lemma:competition_algorithm} proves the correctness of Algorithm~\ref{alg:multi-competition}. \qedhere
            \end{enumerate}
    \end{enumerate}
\end{proof}

\subsection{Agent $a_1$ is a Source, $a_1$ Envies no one in $\B\cup\C$ but $g$-champions \(b_1\)}\label{sec:competition}
In this section, we consider the case when we have a partial EFX allocation $X$ with an unallocated item $g$, in which, agent $a_1$ is a source, and $a_1$ does not envy anyone in $\B\cup \C$
but $a_1$ $g$-champions $b_1$. See the figure below.

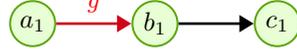
\begin{center}
    \begin{tikzpicture}
        % Nodes
        \node[leading_agent] (a1) {\small \(a_1\) };
        \node[leading_agent, right=1cm of a1] (b1) {\small \(b_1\) };
        \node[leading_agent, right=1cm of b1] (c1) {\small \(c_1\) };
    
        % Edges
        \draw[envy] (b1) -- (c1); 
        \draw[champ] (a1) to[out=0, in=180] node[above, red] {\footnotesize \(g\)} (b1);
    \end{tikzpicture}
\end{center}

The pseudocode for our algorithm to handle this case is presented as Algorithm~\ref{alg:multi-competition}. It takes $X$ as input and outputs another EFX allocation $Y$ such that $\phi(Y)>\phi(X)$.

\subsubsection{Overview of Algorithm~\ref{alg:multi-competition}}
In this section, we give an overview of the algorithm. Please refer to Algorithm~\ref{alg:multi-competition} for the pseudocode. 

\noindent {\bf High-level idea: }Recall that $a_1$ does not envy $b_1$ but $a_1$ $g$-champions $b_1$ with respect to the EFX allocation $X$. So the bundle $T_{b_1}^{a_1}\cup g$ is such that  $a_1$ does not strongly envy anyone after receiving $T_{b_1}^{a_1}\cup g$ and in fact, this hold for any $a_i$, with respect to $X$. Further, $T_{b_1}^{a_1}\cup g>_a X_{a_1}$. Also, no one strongly envies $T_{b_1}^{a_1}\cup g$ with respect to $X$. If $T_{b_1}^{a_1}\cup g$ can be allotted to $a_1$, and a {\em suitable} replacement bundle for $b_1$ can be found, then this will result in an EFX allocation with improved value of $\phi$. This is roughly the goal of Algorithm~\ref{alg:multi-competition}. Temporarily, we allot $T_{b_1}^{a_1}\cup g$ to $b_1$. Clearly, the strong envy can only be from $b_1$ to other agents. This invariant, called Invariant $1$, holds throughout Algorithm~\ref{alg:multi-competition}.

\begin{invariant}\label{inv:only_b1_senvy}
    In the allocation \(X\), only agent \(b_1\) may have strong envy towards others.
\end{invariant}

\noindent {\bf Notation: }At the beginning of Algorithm~\ref{alg:multi-competition}, define $O_b$, the {\em comparison set for $\B$}, as the set of all the bundles allocated to agents in $\A\cup \C$, together with $T_{b_1}^{a_1}\cup g$. During the course of Algorithm~\ref{alg:multi-competition}, some of the bundles in \(O_b\) will be replaced by one of their proper subsets, called a {\em virtual bundle}. At any point in the algorithm, denote $P=\max_b O_b$. Similarly, define $O_c$, the {\em comparison set for $\C$} as the bundles belonging to agents in $\A\cup\B$. Define \(Q=\max_c(O_c \cup \{P\})\).%, the best bundle in \(O_c \cup \{P\}\) according to the agents in \(\C\).

\begin{enumerate}[wide,  labelindent=0pt,label=Case~\arabic*]  
    \item\label{itm:b-likes-a} {\bf $\mathbf{P\in \{X_{a_i}\mid a_i \in \A\}\cup \{T_{b_1}^{a_1}\cup g\}} $:}  See Figure~\ref{fig:fig2}. If $P = X_i$ for some $i$, then $P$ can be allocated to $b_1$ and $T_{b_1}^{a_1}\cup g$ to $a_i$, followed by internal sorting of bundles of groups $\A$ and $\B$. Clearly, after receiving $P$, $b_1$ has no strong envy towards any agent in $\A\cup \C$. Possible strong envy from $b_1$ to other agents in $\B$ can be resolved using Lemma~\ref{lemma:internal_envies} to get an EFX allocation. If $P=X_{a_1}$ then $a_1$ gets a better bundle than $X_{a_1}$, namely $T_{b_1}^{a_1}\cup g$, and hence $\phi$ improves. If $P=X_{a_i}$ for $i\neq 1$, then $X_{a_1}$ still remains the minimum valued bundle in group $\A$. However, $a_1$ envies $P$ since $P=X_{a_i}>_a X_{a_1}$. Thus we can apply Lemma~\ref{lemma:a_1_envies_somebody} to get an EFX allocation with a higher value of $\phi$. A similar argument applies if $P=T_{b_1}^{a_1}\cup g$.

    \begin{figure}[H]
        \centering
        \begin{center}
        \begin{tikzpicture}
            % Nodes
            \node[agent] (bq) {\small \(b_q\)};
            \node[etc, right=0.5cm of bq] (betc) {\(\cdots\)};
            \node[agent, right=0.5cm of betc] (b2) {\small \(b_2\)};
            \node[leading_agent, right=1cm of b2] (b1) {\small \(b_1\)};
            \node[etc, below=0.1cm of b1] (t1_g) {\(T_{b_1}^{a_1}\cup g\)};

            \node[leading_agent, right=2cm of b1] (a1) {\small \(a_1\)};
            \node[etc, right=0.5cm of a1] (a2) {\(\cdots\)};
            \node[agent, right=0.5cm of a2] (ai) {\small \(a_i\)};
            \node[etc, right=0.5cm of ai] (aj) {\(\cdots\)};
            \node[agent, right=0.5cm of aj] (ap) {\small \(a_p\)};

            % % Edges
            \draw[envy] (b1) -- (b2);
            \draw[envy] (b2) -- (betc);
            \draw[envy] (betc) -- (bq);

            \draw[envy] (a1) -- (a2);
            \draw[envy] (a2) -- (ai);
            \draw[envy] (ai) -- (aj);
            \draw[envy] (aj) -- (ap);

            \draw[envy] (a1) -- (b1);
            \draw[best] (b1) to[out=45, in=135] (ai);
            
            % \draw[champ] (c1) to[out=45, in=135] node[below, red] {\footnotesize \(g\)} (b1);
        \end{tikzpicture}
    \end{center}
    \caption{Green edge points to \(P=X_{a_i}=\max_{b}O_b\)}
    \label{fig:fig2}
    \end{figure}

    \item\label{itm:c-likes-a} {\bf $\mathbf{P\subseteq X_{c_i}}$ for some $\mathbf{i}$ and $\mathbf{Q\in \{X_{a_j}\mid a_j \in \A\}\cup \{T_{b_1}^{a_1}\cup g\}} $:} See Figure~\ref{fig:fig3}. If $Q=X_{a_j}$ for some $j$, a similar improvement in $\phi$ can be obtained by giving $T_{b_1}^{a_1}\cup g$ to $a_j$, $P$ to $b_1$, and $Q$ to $c_i$. Note that this might need resolution of intra-group strong envies, using Lemma~\ref{lemma:internal_envies}, and also sorting the bundles internally within each group so that the leading agents retain minimum valued bundles in their respective groups. A similar \(\phi\) improvement can be made if \(Q=T_{b_1}^{a_1}\cup g\) by using Lemma~\ref{lemma:a_1_envies_somebody}.

    \begin{figure}[H]
        \centering
        \begin{center}
        \begin{tikzpicture}
            % Nodes
            \node[agent] (bq) {\small \(b_q\)};
            \node[etc, right=0.5cm of bq] (betc) {\(\cdots\)};
            \node[agent, right=0.5cm of betc] (b2) {\small \(b_2\)};
            \node[leading_agent, right=1cm of b2] (b1) {\small \(b_1\)};
            \node[etc, below=0.1cm of b1] (t1_g) {\(T_{b_1}^{a_1}\cup g\)};

            \node[leading_agent, above=1cm of b1] (c1) {\small \(c_1\)};
            \node[etc, left=0.5cm of c1] (cetc) {\(\cdots\)};
            \node[agent, left=0.5cm of cetc] (ci) {\small \(c_i\)};
            \node[etc, left=0.5cm of ci] (cetc2) {\(\cdots\)};
            \node[agent, left=0.5cm of cetc2] (cr) {\small \(c_r\)};
            
            \node[leading_agent, right=1cm of b1, yshift=0.7cm] (a1) {\small \(a_1\)}; % yshift=1cm
            \node[etc, right=0.5cm of a1] (a2) {\(\cdots\)};
            \node[agent, right=0.5cm of a2] (ai) {\small \(a_j\)};
            \node[etc, right=0.5cm of ai] (aj) {\(\cdots\)};
            \node[agent, right=0.5cm of aj] (ap) {\small \(a_p\)};

            % % Edges
            \draw[envy] (b1) -- (b2);
            \draw[envy] (b2) -- (betc);
            \draw[envy] (betc) -- (bq);

            \draw[envy] (c1) -- (cetc);
            \draw[envy] (cetc) -- (ci);
            \draw[envy] (ci) -- (cetc2);
            \draw[envy] (cetc2) -- (cr);

            \draw[envy] (a1) -- (a2);
            \draw[envy] (a2) -- (ai);
            \draw[envy] (ai) -- (aj);
            \draw[envy] (aj) -- (ap);

            \draw[envy] (a1) to[out=-120, in=20] (b1);
            \draw[best] (b1) to[out=-40, in=140] (ci);
            \draw[best,color=cyan] (ci) to[out=40, in=150] (ai);
            
            % \draw[champ] (c1) to[out=45, in=135] node[below, red] {\footnotesize \(g\)} (b1);
        \end{tikzpicture}
    \end{center}
        \caption{Green edge points to \(P\) and cyan edge to \(Q\)}
        \label{fig:fig3}
    \end{figure}

    \item\label{itm:competition} {\bf Both the above conditions do not hold: } It must be the case that either $P\subseteq X_{c_i}$ and $Q=X_{b_j}$ for some $i,j$, $j\neq 1$ or $Q=P$. However, even in the former case, we cannot simply swap $P$ and $Q$ between $c_i$ and $b_j$, since it is not clear whether this procedure will then terminate. 
    
    \begin{enumerate}[wide,  labelindent=0pt,label*=\alph*]
        \item\label{itm:virtual-update}{\bf  \(\mathbf{Q = X_{b_j}}\) for some \(\mathbf{b_j \in \B \setminus \{b_1\}}\): }
        In this case, the bundle \(X_{c_i}\) in \(O_b\) is replaced by \(X_{c_i} \setminus h\), where \(h\) is the least valued good in \(X_{c_i}\) according to \(v_b\). Note that this operation does not alter the actual allocation. %and therefore no good is deallocated. 
        We refer to this update of \(O_b\) as a \emph{virtual update}. The algorithm starts a new iteration with this updated set $O_b$. The idea is that the virtual update can happen at most once for each bundle of group $\C$. At some point, either $b_1$ chooses such a virtual bundle as $\max_b O_b$, or chooses some bundle of group $\A$, satisfying \ref{itm:b-likes-a} above. 
             %Now, the following observations can be made:

        \item\label{itm:Q=P} {\bf $\mathbf{Q=P=X_{c_i}}$: } Then, agents \(b_1\) and \(c_i\) \emph{compete} for \(P\) with \(O_b\) and \(O_c\) as their respective {\em comparison sets}. Each of $b_1$ and $c_i$ picks the minimum cardinality subsets, say $P_b$ and $P_c$ of $P$ respectively, such that, given this subset, they will have no envy towards any bundle in their respective comparison sets. We call $c_i$ to be the {\em winner} if $|P_c|\leq |P_b|$, and we call $P_c$ as the {\em winning subset}. For a formal definition of the competition framework, please refer to Definition~\ref{def:compitition}.
        
        \begin{enumerate}[wide,  labelindent=0pt,label*=(\roman*)]
            \item\label{itm:c-wins} {\bf \(\mathbf{c_i}\) is a winner and \(\mathbf{P_c \subsetneq X_{c_i}}\):} %In this case, it is possible that both \(c_i\) and \(b_1\) are winners, and the corresponding winning subsets are of equal size. However, the size of these winning subsets is strictly smaller than \(|X_{c_i}|\). 
            We then update the allocation \(X\) by allocating $P_c$ to $c_i$, and adding the goods in \(X_{c_i} \setminus P_c\) to the set \(U\) of unallocated goods. By the definition of a winning subset, when $c_i$ is given $P_c$, agent $c_i$ can have possible strong envies towards only the agents in $\C$. Such strong envies are resolved using Lemma~\ref{lemma:internal_envies}. At this point, we have a new allocation that satisfies Invariant~\ref{inv:only_b1_senvy} with a strictly larger number of unallocated goods. Therefore, the comparison sets \(O_b\) and \(O_c\) are updated to reflect the new allocation: \(O_b \gets \{X_{\alpha} \mid \alpha \in \A \cup \C\} \cup \{T_{b_1}^{a_1} \cup g\}\), and \(O_c \gets \{X_{\alpha} \mid \alpha \in (\A \cup \B)\}\). The algorithm then proceeds with a new iteration. 

            \item\label{itm:same-winning-subset} {\bf \(\mathbf{c_i}\) is a loser, or \(\mathbf{P_c=P_b=X_{c_i}}\).} In this case, a virtual update is done. That is, the bundle \(X_{c_i}\) in \(O_b\) is replaced by \(X_{c_i} \setminus h\), where \(h\) is the least valued good in \(X_{c_i}\) according to \(v_b\). This does not alter the actual allocation \(X\). The algorithm continues with the modified set $O_b$. As in \ref{itm:virtual-update} above, this cannot go on indefinitely.

        \end{enumerate}
        \item\label{itm:virtual-P} {\bf $\mathbf{P}$ is a virtual bundle i.e. $\mathbf{P\subsetneq X_{c_i}}$ for some \(\mathbf{c_i \in \C}\):} This is considered on Line~\ref{line:virtual_bundle} of Algorithm~\ref{alg:multi-competition}. This can happen since $O_b$ might have some virtual bundles from earlier iterations. 
        Thus $P=X_{c_i}\setminus h$. Then either $Q=P$ or $Q$ is a bundle belonging to group $\A$ or $\B$. If $Q=P$, we simply allot $Q$ to $c_i$ and add $h=X_{c_i}\setminus Q$ to the set of unallocated goods $U$. 
        If $Q=X_{b_j}$ for some $j$, we allot $X_{c_i}\setminus h$ to $b_j$, $X_{b_j}$ to $c_i$, and add $h$ to $U$. In both these cases, $h$ gets included in the pool of unallocated items and the pool of unallocated items grows. We update \(O_b\) and \(O_c\) and start a new iteration. Also, it is ensured that if an agent in \(\B\) does not envy a virtual bundle, then she will not \emph{strongly} envy the corresponding real bundle, thus $b_1$ has no strong envy towards agents in $\A\cup\C$. We will prove it in Lemma~\ref{lem:virtual-bundle} later.
    \end{enumerate}
\end{enumerate}

The correctness of \ref{itm:b-likes-a} and \ref{itm:c-likes-a} above is proved respectively in Parts~\ref{lem-part:b-likes-a} and~\ref{lem-part:c-likes-a} of Lemma~\ref{lemma:a_1_envies_none_part_1}.

\begin{definition}\label{def:compitition}
    Let \( a, b \in \N \) be two agents, and let \(O_a\) and \(O_b\) represent the \emph{compare sets} of agents \(a\) and \(b\) respectively, where \( O_a, O_b \subseteq 2^\M \) are collections of bundles of goods. A \emph{competition} occurs when both agents prefer a given bundle \( S \subseteq \M \) over all bundles in their respective compare sets. Formally, agents \(a\) and \(b\) \emph{compete} for \(S\) if:
    \[S >_a T \quad \forall T \in O_a \quad \text{and} \quad S >_b T \quad \forall T \in O_b.\]
    For each agent \(i \in \{a, b\}\), let \( P_i \subseteq S \) denote the smallest subset of \(S\) that agent \(i\) strictly prefers over every bundle in their compare set \(O_i\). The agent with the smaller preferred subset is called the \emph{winner} of the competition. Specifically, agent \(a\) is the \emph{winner} if \( |P_a| \leq |P_b|\), and the \emph{loser} otherwise.
\end{definition}

\begin{algorithm}
\DontPrintSemicolon
\caption{\textsc{Iterated-Competition}}
\label{alg:multi-competition}
\KwIn{An EFX allocation \(X\) such that \(a_1\) \(g\)-champions \(b_1\) and \(a_1\) does not envy \(b_1\)}
\KwOut{An EFX allocation \(Y\) such that \(\phi(Y)>\phi(X)\)}

Allocate  \(\{T_{b_1}^{a_1} \cup g\}\) to agent \(b_1\)\;
Let \(O_b \gets \{X_{\alpha} \mid \alpha \in \A \cup \C \} \cup \{T_{b_1}^{a_1} \cup g\} \) and let \(O_c \gets \{X_{\alpha} \mid \alpha \in (\A \cup \B)\} \)\;
Let \(U\) be the set of unallocated goods in \(X\)\; 

\While{true}{\label{line:while-loop}   
    Let \(P \gets \max_b O_b\)\;
    \If{\(P \in \{ X_{a_i} \mid a_i \in \A \} \cup \{T_{b_1}^{a_1} \cup g\}\)}{
        Apply Lemma~\ref{lemma:a_1_envies_none_part_1} (Part $(i)$), obtaining an EFX allocation \(Y\), such that \(\phi(Y) > \phi(X)\)\;
        \KwRet \(Y\)\;\label{line:b-likes-a}
    }

    \Else(\hfill\tcp*[h]{\(P \subseteq X_{c_i}\) for some \(c_i \in \C\)}){

        Let \(Q \gets \max_c (O_c\cup P)\)\;
        \If{\(Q\in \{ X_{a_i} \mid a_i \in \A \} \cup \{T_{b_1}^{a_1} \cup g\} \)}{
            Apply Lemma~\ref{lemma:a_1_envies_none_part_1} (Part $(ii)$), obtaining an EFX allocation \(Y\), such that \(\phi(Y) > \phi(X)\)\;
            \KwRet \(Y\)\;\label{line:c-likes-a}
        }

        \ElseIf{\(P = X_{c_i}\) for some \(c_i \in \C\)}{
            \If{\(Q=P\)}{
                \(b_1\) and \(c_i\) \emph{compete} for \(P\) \;%with \(O_b\) and \(O_c\cup \{P\}\) as their respective compare sets.\;
                \If{\(c_i\) is a winner \textbf{and} the winning subset \(P_c \subsetneq P\)}{
                    In \(X\), replace \(X_{c_i}\) with the winning subset \(P_c\subsetneq X_{c_i}\)\;
                    \(U\gets U\cup (X_{c_i} \setminus P)\)\;
                    Resolve internal strong envies in \(\C\) (Lemma~\ref{lemma:internal_envies})\;
                    Update: \(O_c \gets \{X_{\alpha} \mid \alpha \in (\A \cup \B)\} \) and \(O_b \gets \{X_{\alpha} \mid \alpha \in \A \cup \C \} \cup \{T_{b_1}^{a_1} \cup g\} \)\;
                }
                \Else{
                    In \(O_b\), replace \(X_{c_i}\) with \(X_{c_i} \setminus h\), where \(h = \min_b X_{c_i}\)\;
                }
            }
            \ElseIf{\(Q\in \{ X_{b_j} \mid b_j \in \B \setminus \{b_1\} \}\)}{
                In \(O_b\), replace \(X_{c_i}\) with \(X_{c_i} \setminus h\), where \(h = \min_b X_{c_i}\)\;
            }
            
        }
        \ElseIf{\(P \subsetneq X_{c_i}\) for some \(c_i \in \C\)}{\label{line:virtual_bundle}
          \If{\(Q=P\)}{
                In \(X\), replace \(X_{c_i}\) with \(P\)\;  
            }

          \Else(\hfill \tcp*[h]{\(Q = X_{b_j}\) for some \(b_j \in  \B \setminus \{b_1\}\)}){

                {\bf Modify \(X\):} allocate \(P\) to \(b_j\), \(X_{b_j}\) to \(c_i\)\;\label{line:modify}
                Resolve \(b_j\) to \(b_k\in \B\setminus b_1\) strong envies by replacing \(X_{b_k}\) with \(\mes_b(X_{b_k},P)\);\

            Resolve internal strong envies in \(\C\) (Lemma~\ref{lemma:internal_envies})\;
            }
        \(U\gets U\cup (X_{c_i} \setminus P)\)\;
        Update: \(O_b \gets \{X_{\alpha} \mid \alpha \in \A \cup \C \} \cup \{T_{b_1}^{a_1} \cup g\} \) and \(O_c \gets \{X_{\alpha} \mid \alpha \in (\A \cup \B)\} \)\;
        }
    }
}
\end{algorithm}

\begin{lemma}\label{lemma:a_1_envies_none_part_1}
    Let \(X\) be a partial allocation with an unallocated good \(g\), where all strong envies in \(X\), if any, originate from agent \(b_1\). Suppose agent \(a_1\) is a source in \(E_X\), does not envy \(b_1\), but \(g\)-champions \(b_1\), partitioning \(X_{b_1}\) as \(X_{b_1} = T_{b_1}^{a_1} \uplus B_{b_1}^{a_1}\). Define:
    \begin{align*}
        O_b &= \{X_{\alpha} \mid \alpha \in \A \cup \C\} \cup \{T_{b_1}^{a_1} \cup g\}\\
        O_c &= \{X_{\alpha} \mid \alpha \in (\A\cup \B) \setminus \{b_1\}\} \cup \{T_{b_1}^{a_1}\cup g\}
    \end{align*}
    If either of the following conditions is satisfied, then there exists an EFX allocation \(Y\) (possibly partial) such that \(\phi(Y) > \phi(X)\).
    
    \begin{enumerate}[leftmargin=*]
        \item\label{lem-part:b-likes-a} There exists a bundle \(P\in \{X_{\alpha} \mid \alpha \in \A \} \cup \{T_{b_1}^{a_1} \cup g\} \) such that, given \(P\), agent \(b_1\) does not strongly envy any bundle in \(O_b\),
        
        \item\label{lem-part:c-likes-a} There exists a bundle \(P\subseteq X_{c_i}\) for some \(c_i \in \C\), such that given \(P\), agent \(b_1\) does not strongly envy any bundle in \(O_b\), and there exists some bundle \(Q\in \{X_{\alpha} \mid \alpha \in \A \} \cup \{T_{b_1}^{a_1} \cup g\}\) such that with \(Q\) agent \(c_i\) does not strongly envy any bundle in \(O_c\cup \{P\}\).
        
    \end{enumerate}
    
\end{lemma}

\begin{proof}
\begin{enumerate}[wide,  labelindent=0pt, label=(\roman*)]
    \item\label{itm:S=Tuniong} If \(P=\{T_{b_1}^{a_1} \cup g\}\), then we allocate \(\{T_{b_1}^{a_1} \cup g\}\) to \(b_1\). We know that with \(P\), agent \(b_1\) does not strongly envy anyone from \(\A\cup\C\). Resolve internal envies in \(\B\), if any, using Lemma~\ref{lemma:internal_envies}. We now have an EFX allocation \(Y\) such that agent \(a_1\) envies \(b_1\). Therefore, by applying Lemma~\ref{lemma:a_1_envies_somebody} on \(Y\), we get an EFX allocation \(Z\) such that \(\phi(Z)>\phi(X)\).
    
    If \(P=X_{a_1}\), we then give \(P\) to agent \(b_1\) and \(\{T_{b_1}^{a_1} \cup g\}\) to agent \(a_1\). Resolve internal envies in \(\B\), if any, using Lemma~\ref{lemma:internal_envies}. We now have an EFX allocation \(Y\) such that \(\phi(Y)>\phi(X)\). 

    Finally, if \(P=X_{a_i}\) for some \(a_i \in \A\setminus a_1\), then give \(P\) to agent \(b_1\) and \(\{T_{b_1}^{a_1} \cup g\}\) to agent \(a_i\). Resolve internal envies in \(\B\) or \(\A\), if any, using Lemma~\ref{lemma:internal_envies}.  This gives an EFX allocation \(Y\) such that agent \(a_1\) envies \(b_1\). Therefore, by applying Lemma~\ref{lemma:a_1_envies_somebody} on \(Y\), we get an EFX allocation \(Z\) such that \(\phi(Z)>\phi(X)\).

    \item\label{itm:S=Xci} Let \(P\subseteq X_{c_i}\) for some agent \(c_i\in \C\). If \(Q=\{T_{b_1}^{a_1} \cup g\}\), we allocate \(P\) to \(b_1\) and \(Q\) to agent \(c_i\). Else if \(Q=X_{a_j}\) for some \(a_j\in \A\setminus a_1\) we allocate \(P\) to \(b_1\), \(Q\) to \(c_i\) and \(\{T_{b_1}^{a_1} \cup g\}\) to agent \(a_j\). In either case, we resolve internal envies in within the groups, if any, using Lemma~\ref{lemma:internal_envies}. We now have a EFX allocation \(Y\) such that agent \(a_1\) envies \(c_i\). Therefore, by applying Lemma~\ref{lemma:a_1_envies_somebody} on \(Y\), we get an EFX allocation \(Z\) such that \(\phi(Z)>\phi(X)\).

    If \(Q=X_{a_1}\), then we allocate \(P\) to \(b_1\),\(Q\) to \(c_i\), and \(\{T_{b_1}^{a_1} \cup g\}\) to agent \(a_1\). Resolve internal envies in within the groups, if any, using Lemma~\ref{lemma:internal_envies}. We get an EFX allocation \(Y\) such that \(\phi(Y)>\phi(X)\). \qedhere
    \end{enumerate} 
\end{proof}

\begin{lemma}\label{lem:virtual-bundle}
    In Algorithm~\ref{alg:multi-competition}, if $b_1$ is allocated a virtual bundle from $O_b$, then $b_1$ does not strongly envy anyone from $\A\cup\C$. In fact, the same holds for any $b_j\in \B$.
\end{lemma}
\begin{proof}
 A virtual bundle appears in $O_b$ when \ref{itm:virtual-update} holds i.e., $\max_b O_b=X_{c_i}$ for some $i$ and $\max_c O_c=X_{b_j}$ for some $j\neq 1$, or \ref{itm:same-winning-subset} holds, when $P=Q=X_{c_i}$ and either $b_1$ has a smaller winning subset than $c_i$ i.e. $|P_b|<|P_c|$ or both have $X_{c_i}$ as the winning subset i.e., $P_b=P_c=X_{c_i}$. 
When $P=\max_b O_b$ is a virtual bundle, then for any non-virtual i.e., {\em real} bundle $S\in O_b$, $P>_b S$ and hence $b_1$ does not envy the agent holding $S$. For any virtual bundle $S=X_{c_i}\setminus h\in O_b$, $S\neq P$, we have $P>_bX_{c_i}\setminus h$ for minimum valued item $h\in X_{c_i}$. Hence $b_1$ might envy the real bundle $X_{c_i}$ but it is not a strong envy. As $b_j$ and $b_1$ have the same valuation $v_b$, the lemma holds for any $b_j\in B$.\qedhere
\end{proof}

To prove the correctness of \ref{itm:competition}, we show that Invariant~\ref{inv:only_b1_senvy} is maintained as long as Algorithm~\ref{alg:multi-competition} remains in \ref{itm:competition}, and that the algorithm exits from \ref{itm:competition} in a finite number of iterations.
\begin{lemma}\label{lem:competition-correctness}
At any point during the course of Algorithm~\ref{alg:multi-competition}, if \ref{itm:competition} holds, then in a finite number of iterations, the algorithm exits from \ref{itm:competition}, and further, Invariant~\ref{inv:only_b1_senvy} is maintained as long as the algorithm remains in \ref{itm:competition}. 
\end{lemma}
\begin{proof}
    In each iteration of Algorithm~\ref{alg:multi-competition}, whenever \ref{itm:competition} holds, either the pool of unallocated items increases or the number of virtual bundles in $O_b$ grows. The former happens in cases \ref{itm:c-wins} and \ref{itm:virtual-P} whereas the later happens in cases \ref{itm:virtual-update} and \ref{itm:same-winning-subset}. Since neither of these can continue beyond a finite number of steps (at most as many steps as the number of items), Algorithm~\ref{alg:multi-competition} exits \ref{itm:competition} in a finite number of steps. 

    It can be seen that Invariant~\ref{inv:only_b1_senvy} holds as long as Algorithm~\ref{alg:multi-competition} continues in \ref{itm:competition}. When the allocation is modified in \ref{itm:competition} (Line~\ref{line:modify} of Algorithm~\ref{alg:multi-competition}), $b_j,j\neq 1$ gets a virtual bundle $P=X_{c_i}\setminus h$. By Lemma~\ref{lem:virtual-bundle}, on allocating $P$, $b_j$ does not strongly envy anyone from $\A\cup\C$. Similarly, by the choice of $Q$, $c_i$ does not strongly envy anyone from $\A\cup \C$. No agent except possibly $b_1$ strongly envies $b_j$ since they did not strongly envy $c_i$ when $c_i$ had $X_{c_i}$. The potential strong envies from $b_j$ towards other agents in $\B$ and from $c_i$ to other agents in $\C$ are resolved using Lemma~\ref{lemma:internal_envies}. Thus only $b_1$ might strongly envy, and hence Invariant~\ref{inv:only_b1_senvy} continues to hold.\qedhere
    
\end{proof}

The following lemma completes the correctness proof of Algorithm~\ref{alg:multi-competition}.
\begin{lemma}\label{lemma:competition_algorithm}
    Let \(X\) be a partial EFX allocation with at least one unallocated good \(g\). Let agent \(a_1\) be a source in \(E_X\) and does not envy agent \(b_1\), but \(g\)-champions \(b_1\). Then, with \(X\) as input, Algorithm~\ref{alg:multi-competition} terminates with an EFX allocation \(Y\) such that \(\phi(Y)>\phi(X)\).
\end{lemma}

\begin{proof}
    If \ref{itm:b-likes-a} holds at any point, the algorithm, on Line~\ref{line:b-likes-a} returns an EFX allocation $Y$ with $\phi(Y)>\phi(X)$ by Part~\ref{lem-part:b-likes-a} of Lemma~\ref{lemma:a_1_envies_none_part_1}. If \ref{itm:c-likes-a} holds, then on Line~\ref{line:c-likes-a}, an EFX allocation $Y$ with $\phi(Y)>\phi(X)$ is returned by Part~\ref{lem-part:c-likes-a} of Lemma~\ref{lemma:a_1_envies_none_part_1}. So the algorithm successfully terminates whenever one of these two cases hold.

    Note that the algorithm terminates if and only if it reaches Line~\ref{line:b-likes-a} or Line~\ref{line:c-likes-a} i.e. if and only if \ref{itm:b-likes-a} or \ref{itm:c-likes-a} hold. So the while loop on Line~\ref{line:while-loop} can go on as long as the algorithm continues in \ref{itm:competition}. As proved in Lemma~\ref{lem:competition-correctness}, \ref{itm:competition} can hold for only a finite number of iterations. So the algorithm must enter \ref{itm:b-likes-a} or \ref{itm:c-likes-a} in a finite number of iterations, and must terminate with an EFX allocation $Y$ such that $\phi(Y)>\phi(X)$.\qedhere
\end{proof}

From Lemma~\ref{lemma:a_1_envies_somebody} and Lemma~\ref{lemma:a_1_envies_none}, we derive the following corollary.
\begin{corollary}\label{corollary:a_1_is_a_source}
    Let \(X\) be a partial EFX allocation with at least one unallocated good \(g\). If \(a_1\) is a source in \(E_X\), then there exists an EFX allocation \(Y\) such \(\phi(Y)>\phi(X)\).
\end{corollary}

\section{Agent $a_1$ is Not a Source in $E_X$}\label{sec:not_a_source}
When $a_1$ is not a source in $E_X$, someone must envy $a_1$. By Proposition~\ref{prop:M_X_is_cyclic}, one of $b_1$ or $c_1$ must envy $a_1$. In addition to this, $a_1$ may or may not envy anyone from $\B\cup\C$. In Lemma~\ref{lemma:a_1_is_not_a_source_easy}, we consider the case when $a_1$ envies someone from $\B\cup\C$. The other case when $a_1$ does not envy anyone from $\B\cup\C$ is considered in Lemma~\ref{lemma:a_1_is_not_a_source_hard}. In each of these cases, we give a construction of another EFX allocation that has a higher value of the potential function.

\begin{lemma}\label{lemma:a_1_is_not_a_source_easy}
    Let \(X\) be a partial EFX allocation with at least one unallocated good \(g\). If \(a_1\) is a not a source in \(E_X\), and if \(a_1\) envies some agent in \(\B\cup\C\), then there exists an EFX allocation \(Y\) such that \(\phi(Y)>\phi(X)\).
\end{lemma}
\begin{proof}
    since $a_1$ is not a source, either \(b_1\) or \(c_1\) must envy \(a_1\) by Proposition~\ref{prop:M_X_is_cyclic}. Without loss of generality, assume \(b_1\) envies \(a_1\). We know that \(a_1\) envies some agent from \(\B\cup \C\). We proceed by considering the cases based on which agent is envied.
    \begin{enumerate}[wide,  labelindent=0pt, label=Case~\arabic*]
        \item\label{itm:a1-envies-bj} {\bf Agent \(a_1\) envies some agent \(b_j\in \B\):} Then, \(\langle(a_1,b_j),(b_1,a_1) \rangle\) is a pseudo-cycle. Therefore, using Lemma~\ref{lemma:pseudo-cycle} we get a new EFX allocation \(Y\) such that \(\phi(Y)>\phi(X)\).
    
    \item\label{itm:a1-envies-cj} {\bf Agent \(a_1\) envies some \(c_j\in \C\):} If \(c_1\) envies \(a_1\), then \(\langle (a_1,c_j),(c_1,a_1)\rangle\) is a pseudo-cycle. Thus we are done. Consider the case when \(c_1\) does not envy \(a_1\). We know that some leading agent must \(g\)-champion \(b_1\). We consider all possible cases based on who \(g\)-champions \(b_1\).
        \begin{enumerate}[wide,  labelindent=0pt, label*=.\arabic*]
        \item\label{itm:b1-self-champion} {\bf Agent \(b_1\) self \(g\)-champions:} We make a Pareto improvement by replacing \(X_{b_1}\) with \(\mes_b(X_{b_1} \cup g, X_{b_1})\). Following this, the bundles in \(\B\) are re-sorted so that \(b_1\) holds the least valuable bundle. Call this allocation \(Y\). If in \(Y\), \(b_1\) does not envy \(a_1\), then \(a_1\) is a source in \(E_Y\). Therefore, applying Corollary~\ref{corollary:a_1_is_a_source}, we get a new EFX allocation \(Z\) such that \(\phi(Z)>\phi(Y)=\phi(X)\). Otherwise, if \(b_1\) envies \(a_1\) in \(Y\), then we apply Lemma~\ref{lemma:a_1_is_not_a_source_easy} again on \(Y\). As there are finitely many EFX allocations, this Pareto improvement cannot go on forever.

        \item\label{a1-g-champions-b1} {\bf Agent \(a_1\) \(g\)-champions \(b_1\):} Then \(\langle(a_1,b_1),(b_1,a_1)\rangle\) is a pseudo-cycle. Therefore, applying Lemma~\ref{lemma:pseudo-cycle},  we get new EFX allocation \(Y\) such that \(\phi(Y)>\phi(X)\).
        
        \item\label{c1-g-champions-b1} {\bf Agent \(c_1\) \(g\)-champions \(b_1\):} Then we have \(\langle (a_1,c_j),(c_1,b_1),(b_1,a_1)\rangle\) as  a pseudo-cycle. Thus, we get new EFX allocation \(Y\) such that \(\phi(Y)>\phi(X)\).\qedhere
        \end{enumerate}
        \end{enumerate}
\end{proof}

\begin{restatable}{lemma}{anotsourcehard}\label{lemma:a_1_is_not_a_source_hard}
    Let \(X\) be a partial EFX allocation with at least one unallocated good \(g\). If \(a_1\) is a not a source in \(E_X\), and if \(a_1\) envies no agent in \(\B\cup\C\), then there exists an EFX allocation \(Y\) such that either \(Y\succ X\) or \(\phi(Y)>\phi(X)\).
\end{restatable}
\begin{proof}
    We know that either \(b_1\) or \(c_1\) must envy \(a_1\). Without loss of generality, assume \(b_1\) envies \(a_1\). Some leading agent must \(g\)-champion \(b_1\). If \(b_1\) self \(g\)-champions, then we can make a Pareto improvement, thus we are done. If \(a_1\) \(g\)-champions \(b_1\), we have a pseudo-cycle \(\langle (a_1,b_1),(b_1,a_1)\rangle\). Therefore, consider the case when agent \(c_1\) \(g\)-champions \(b_1\). 

    We know that some leading agent must \(g\)-champion \(c_1\). If \(c_1\) self \(g\)-champions, then we can make a Pareto improvement. Suppose \(b_1\)  \(g\)-champions \(c_1\). If either \(c_1\) envies \(b_1\) or \(b_1\) envies \(c_1\), then \((c_1,b_1,c_1)\) is a Pareto improving cycle, thus we make a Pareto improvement. If not, we have \(b_1\) and \(c_1\) \(g\)-championing each other and \(a_1\) envies neither \(b_1\) nor \(c_1\). We apply Lemma~\ref{lemma:mx_two_cycle} to get Pareto improvement. 

    The only case remaining is when agent \(a_1\) \(g\)-champions \(c_1\). In this case, we ask \(c_1\) to pick the best bundle from agents in \(\A \cup \B\).
    \begin{enumerate}[wide,  labelindent=0pt, label=Case~\arabic*]
        \item\label{itm:c1-picks-bj} {\bf Agent \(c_1\) picks a bundle \(X_{b_j}\) from an agent \(b_j\in \B\):} Then, we construct a new allocation \(Y\), by allocating \(X_{b_j}\) to \(c_1\), \(T_{c_1}^{a_1}\cup g\) to agent \(a_1\), and \(X_{a_1}\) to \(b_j\). Here, agent \(a_1\) strictly improves. In this new allocation, no agent strongly envies any of  \(a_1,b_j\) and \(c_1\) as the bundles that these agents received are either \(\mes\) bundles or complete bundles in the old allocation \(X\). Agent \(b_j\) does not strongly envy any agent as \(v_b(Y_{b_j})\ge v_b(Y_{b_1})\) and agent \(b_1\) does not strongly envy any agent. As \(a_1\) received a strict improvement, she does not strongly envy any agent. Finally, we know that \(c_1\) does not envy any agent from \(\B\cup \A\) as she chose the best bundle. If \(c_1\) strongly envies any agent in \(\C\), we resolve them using Lemma~\ref{lemma:internal_envies}. We now have \(\phi(Y)>\phi(X)\).

        \item\label{itm:c1-picks-a1} {\bf Agent \(c_1\) picks a bundle \(X_{a_1}\):} Then, to construct a new allocation \(Y\), allocate \(X_{a_1}\) to \(c_1\) and \(T_{c_1}^{a_1}\cup g\) to \(a_1\). Again, note that if there are any strong envies in \(Y\), they must be from agent \(c_1\) to other agents in \(\C\). We resolve such strong envies using Lemma~\ref{lemma:internal_envies}. We have \(\phi(Y)>\phi(X)\).

        \item\label{itm:c1-picks-aj} {\bf Agent \(c_1\) picks \(X_{a_j}\), \(a_j\in \A\setminus \{a_1\}\):} Then, to construct a new allocation \(Y\), allocate \(X_{a_j}\) to \(c_1\) and \(T_{c_1}^{a_1}\cup g\) to \(a_j\). Note that agent \(a_j\) does not strongly envy anyone as \(v_a(Y_{a_j})>v_a(Y_{a_1})= v_a(X_{a_1})\) and agent \(a_1\) does not strongly anyone. If there are any strong envies in \(Y\), they must be from agent \(c_1\) to other agents in \(\C\). We resolve such strong envies using Lemma~\ref{lemma:internal_envies}. In \(Y\), agent \(b_1\) continues to envy \(a_1\) as their bundles did not change. Agent \(a_1\) envies \(c_1\) as \(a_1\) previously envied \(a_j\). Therefore, we apply Lemma~\ref{lemma:a_1_is_not_a_source_easy} on \(Y\) to get a new allocation \(Z\) such that \(\phi(Z)>\phi(Y)=\phi(X)\).\qedhere
    \end{enumerate}
\end{proof}

From Lemma~\ref{lemma:a_1_is_not_a_source_easy} and Lemma~\ref{lemma:a_1_is_not_a_source_hard}, we get the following corollary.

\begin{corollary}\label{corollary:a_1_not_a_source}
    Let \(X\) be a partial EFX allocation with at least one unallocated good \(g\). If \(a_1\) is a not a source in \(E_X\), then there exists an EFX allocation \(Y\) such either \(Y\succ X\) or \(\phi(Y)>\phi(X)\).
\end{corollary}

Therefore, Corollary~\ref{corollary:a_1_is_a_source} and Corollary~\ref{corollary:a_1_not_a_source} complete the proof of Theorem~\ref{thm:efx3}. Hence, EFX exists for three types of agents. 

\section{Conclusion}\label{sec:conclusion}
In this paper, we proved the existence of EFX allocations for instances with three types of agents, where agents of the same type have identical additive valuations. Our result extends prior work on EFX allocations for three agents \cite{efx_3}, and for two types of agents \cite{mahara}.

The techniques developed in this work may provide insights into more general settings. A natural direction for future research is to extend these results to agents with more general valuation functions, beyond the additive case. Another long-standing open question is whether EFX allocations can be guaranteed for instances involving four or more agents. 

Given the progress made here, we propose the following conjecture as a potential pathway for further exploration:
\begin{conjecture}
If EFX allocations exist for \(k\) agents with additive valuations, for some constant \(k\), then EFX allocations exist for any number of agents with at most \(k\) distinct types.
\end{conjecture}

This conjecture, if proven, would broaden the scope of EFX allocations, particularly in practical settings where large groups of agents share similar preferences.

\section*{Acknowledgements}
We are deeply grateful to the anonymous reviewers of ACM EC 2025 for their exceptionally thorough and thoughtful reviews. Their detailed feedback went far beyond the norm and significantly improved the clarity and quality of this work. Vishwa Prakash HV acknowledges support from the TCS-RSP Research Fellowship. 

\bibliographystyle{alphaurl}%alpha}%{plainnat} 
\bibliography{references}

\end{document}